\documentclass[aps,prb,twocolumn,showpacs,superscriptaddress,groupedaddress,floatfix,longbibliography]{revtex4-2}
\usepackage{mathtools}
\usepackage{latexsym}
\usepackage{graphicx}
\usepackage{float}
\usepackage{dcolumn}
\usepackage{bm}
\usepackage{amssymb}
\usepackage{amsmath}
\usepackage{mathtools}
\usepackage[space]{grffile}
\usepackage{xcolor}

\newcommand{\DeltaE}{E_0}

\begin{document}

\title{Non-integrable Floquet Ising model with duality twisted boundary conditions}
\author{Aditi Mitra$^{1,3}$} \author{Hsiu-Chung Yeh$^{1}$} \author{Fei Yan$^{2}$}
\author{Achim Rosch$^{3}$}
\affiliation{$^{1}$Center for Quantum
  Phenomena, Department of Physics, New York University, 726 Broadway,
  New York, NY, 10003, USA \\$^{2}$NHETC and Department of Physics and Astronomy, Rutgers University,
136 Frelinghuysen Rd, Piscataway, NJ 08854, USA\\$^{3}$Institute for Theoretical Physics, University of Cologne, 50937 Cologne, Germany
  }
\begin{abstract}
Results are presented for a Floquet Ising chain with duality twisted boundary conditions, taking into account the role of weak integrability breaking in the form of four-fermion interactions. In the integrable case, a single isolated Majorana zero mode exists which  
is a symmetry in the sense that  it commutes both with the Floquet unitary and the $Z_2$ symmetry of the Floquet unitary. When integrability is weakly broken, both in a manner so as to preserve or break the $Z_2$ symmetry, the Majorana zero mode is still found to be conserved for small system sizes. This is reflected in the dynamics of an infinite temperature autocorrelation function which, after an initial transient that is controlled by the strength of the integrability breaking term, approaches a plateau 
that does not decay with time. The height of the plateau agrees with a numerically constructed conserved quantity, and is found to decrease with increasing system sizes. 
It is argued that  the existence of the plateau and its vanishing for larger system sizes is closely related to a localization-delocalization transition in Fock space triggered by the integrability-breaking interactions.
\end{abstract}
\maketitle

\section{Introduction}

The transverse field Ising model is one of the most basic models for understanding fundamental concepts in condensed matter physics such as phase transitions, topological order, non-Abelian excitations and duality. The famous Kramers-Wannier duality transformation \cite{Kramers1941} maps the Ising model to its dual, where the dual is also an Ising model
but with the couplings corresponding to the Ising exchange interactions and the transverse fields, interchanged. At the level of wavefunctions, the duality transformation maps the doubly degenerate ferromagnetic ground state to the non-degenerate paramagnetic ground state. Since this is a two to one mapping, the duality transformation, when represented as an operator, is
a non-invertible operator \cite{Fendley16,Fendley20}.

The duality transformation can be implemented through a duality defect, which has some remarkable topological properties, namely that it locally commutes with the transfer matrix of the corresponding two-dimensional classical model or the generator of time-evolution in the quantum 1+1D model \cite{Fendley16, Fendley20, Verstraete21}. In fact, the  transverse field Ising model 
hosts two topological defects, one is related to the $Z_2$ or spin flip symmetry (${\mathcal D}_{\psi}$) and the other performs the duality transformation ($\mathcal{D}_{\sigma}$). These topological defects obey the Ising anyon fusion rules $\mathcal{D}_{\sigma}^2 = \mathcal{I} + \mathcal{D}_{\psi}, \mathcal{D}_{\psi}\mathcal{D}_{\sigma}= \mathcal{D}_{\sigma}\mathcal{D}_{\psi}= \mathcal{D}_{\sigma}, \mathcal{D}_{\psi}^2=\mathcal{I}$, with $\mathcal{I}$ representing the identity operator. These topological defects, while traditionally studied in the context of ground state properties and/or in the context of conformal field theories \cite{Verlinde:1988sn,CARDY1989581,Petkova00,FrohlichFucksRunkelSchweigert,Chang:2018iay}, can in fact be constructed on the lattice \cite{Fendley16,Fendley20}. Moreover, this construction can also be generalized to the Floquet version of the Ising model \cite{Tan22}. It has also been shown how to implement the duality transformation using gates and measurements \cite{Verresen21, Aasen_2022}, where depending on the measurement outcome, four different kinds of Kramers-Wannier duality transformations may be identified \cite{Aasen_2022}. 

Topological defects can also generate non-trivial boundary conditions in the system \cite{Levy91,Schutz93,Oshikawa97,Petkova00,Pearce01,Grimm:2001dr,Fendley16,Saleur21}. Consider a (1+1)-D system placed on a spatially circular chain, a time-like spin-flip defect corresponds to imposing anti-periodic boundary conditions along the circular chain.  
Similarly a time-like duality defect corresponds to imposing  duality twisted boundary conditions. Recently, the integrable Floquet Ising model with duality twisted boundary conditions was studied \cite{Tan22}. It was shown that the duality twist 
allows the chain to host a single isolated Majorana zero  mode (rather than a pair of Majorana zero modes), despite the Floquet driving. This mode was detected in two ways. The first was by the full analytic construction of the Majorana operator,
and the second was by numerically studying the autocorrelation function of a local operator that has an overlap with the Majorana mode. It is noteworthy that the Majorana zero mode that exists in the presence of a duality twist does not lead to any
degeneracy or pairing of the eigenspectra of the Floquet unitary, 
in contrast to Majorana modes (also known as strong modes) that appear with open boundary conditions
\cite{FendleyXYZ, Sen13, Yates19}.
In this paper we discuss the effects of weak integrability breaking terms on the duality twisted Floquet Ising chain. 

The paper is organized as follows. In Section \ref{sec:Int} we review some of the properties of the integrable Floquet Ising model with a duality twist. We also present new results such as the effect of translating the duality twist along the integrable Floquet chain, and we highlight some subtleties of the dynamics of the Majorana operators. In Section \ref{sec:NInt} we present results for the non-integrable Floquet Ising model, showing how the Majorana mode is still remarkably stable for finite system sizes. In Section \ref{sec:Pl} we present a phenomenological argument for the stability of the Majorana zero mode. We present our conclusions in Section \ref{sec:Conc}.

\section{Integrable Floquet Ising model with a duality twist} \label{sec:Int}
Let us consider a
chain with $2L$ sites labeled by $0,1,2 \ldots 2L-1$. The spins reside on the odd sites $1,3, \ldots 2L-1$, while the even sites are empty (and can be considered to be dual sites) \cite{Fendley16,Tan22}. 
We impose periodic boundary conditions where site $n$ is the same as site $2L+n$. We denote $X_j, Y_j, Z_j$ as the Pauli operators on site $j$.
The usual defectless and integrable Floquet Ising unitary, allowing for spatially 
inhomogeneous couplings is
\begin{align}
U = \left(\prod_{j=0}^{L-1}W_{2j+1}^{X}(u_{2j+1})\right)\left(\prod_{j=0}^{L-1}W_{2j}^{ZZ}(u_{2j})\right),
\end{align}
where
\begin{subequations}
  \begin{align}
    W_{2j+1}^X(u_{2j+1}) &= e^{-i u_{2j+1}X_{2j+1}},\\
    W_{2j}^{ZZ}(u_{2j}) &= e^{-i u_{2j} Z_{2j-1}Z_{2j+1}}.
\end{align}  
\end{subequations}

Above, $W^X_{2j+1}(u_{2j+1})$ represents the single-site unitary gate that corresponds to applying a transverse field of strength $u_{2j+1}$ along $X$ on the spin at site $2j+1$. $W^{ZZ}_j(u_{2j)}$ is the two-site unitary gate that corresponds to a $ZZ$ type Ising interaction of strength $u_{2j}$ between spins at sites $2j-1$ and $2j+1$. 
The above unitary has a discrete $Z_2$
symmetry generated by
\begin{align}
    D_{\psi} = X_1 X_3 \ldots X_{2L-1},
\end{align}
which describes the simultaneous flip of all spins. While for the  defectless case presented above, $u_{\rm even}$ corresponds to Ising interactions while
$u_{\rm odd}$ corresponds to the strength of the transverse field, we will 
show later that a duality defect will interchange the role of the two couplings. 

\begin{figure}
\includegraphics[width = 0.45\textwidth]{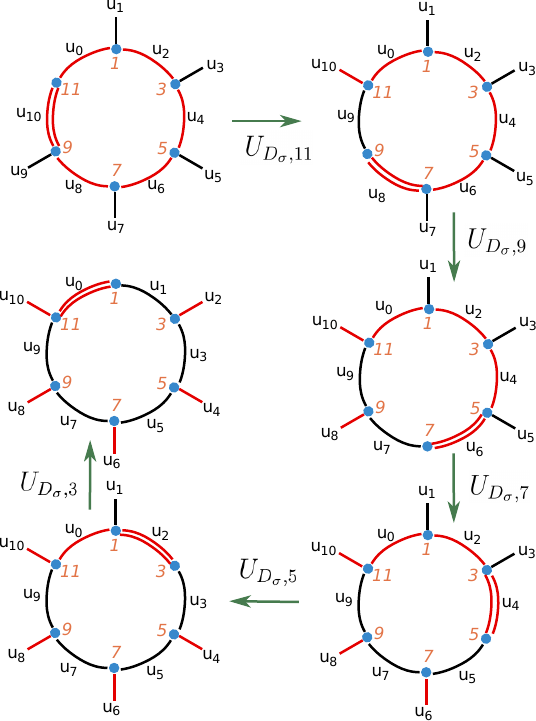}
\caption{Duality twist and its translation for $L=6$. Top left: Initial configuration where a duality twist (double red line)  exists between sites $9$ and $11$, where the mixed coupling term is $u_{10}Z_{9}X_{11}$. All the odd couplings (black) correspond to transverse fields ($u_{2j+1}X_{2j+1},j\neq 5$), while all the even couplings (red) are Ising exchange interactions ($u_{2j}Z_{2j-1}Z_{2j+1}, j\neq 5$). A Majorana zero mode is localized around site $11$. Top right: The twist has been moved to the neighboring link, i.e, between sites $7$ and $9$. In this process, a domain wall is introduced at site $11$ where $u_9, u_{10}$ exchange roles. The Majorana continues to reside around site $11$. Other panels: Further translations of the duality twist do not move the Majorana zero mode. Meanwhile the Kramers-Wannier transformation is applied to the region between the duality twist and the domain wall, interchanging the roles of $u_{\rm even}$ and $u_{\rm odd}$. 
}\label{shift}
\end{figure}

The duality twisted boundary conditions on the $2L-2$ link, i.e, the link between sites $2L-3, 2L-1$ corresponds to \cite{Fendley16,Tan22} removing the transverse field at site $2L-1$, and changing the Ising coupling between sites $2L-3$ and $2L-1$  from $Z_{2L-3}Z_{2L-1}$ to a mixed Ising coupling, $Z_{2L-3}X_{2L-1}$. This is shown in the top left panel of Fig.~\ref{shift} for  $L=6$.  Thus the Floquet unitary with a duality twist between sites $2L-3$ and $2L-1$ is
\begin{align}
 T_{\sigma, 2L-1}= &\left(\prod_{j=0}^{L-2}W_{2j+1}^{X}(u_{2j+1})
 \right)  e^{-iu_{2L-2}Z_{2L-3}X_{2L-1}}   \nonumber\\
 &\times \left(\prod_{j=0}^{L-2}
 W_{2j}^{ZZ}(u_{2j})\right). \label{Tsig}
\end{align}
In particular, $u_{2L-2}$ controls the strength of the exchange interaction, albeit of the mixed coupling type $ZX$. Moreover, $u_{2L-1}$ is missing due to the absence of the transverse field on site $2L-1$. The symmetry operator in the presence of the duality twist is no longer $D_{\psi}$ but $\Omega$ where 
\begin{align}
    \Omega &= i Z_{2L-1} D_\psi = -(X_1\ldots X_{2L-3})Y_{2L-1}.\label{Omdef}
\end{align}

We will show later that unitary transformations that shift the location of the duality defect, will locally (along the trajectory of the shifted defect) interchange the  role of the couplings $u_{\rm even}(u_{\rm odd})$ from an Ising (transverse field) coupling, to a transverse field (Ising) coupling.

The duality twisted boundary condition is a highly non-local perturbation in the language of Majorana fermions. In order to show this, we  rewrite the Floquet unitary in terms of the Majorana fermions  
\begin{subequations}\label{JW}
\begin{align}
    \gamma_{2j} &= \left(\prod_{k=-1}^{j-1}X_{2k+1}\right)Z_{2j+1},\,\, j= 0,\ldots,L-2,\\ 
\gamma_{2j+1} &=- \left(\prod_{k=-1}^{j-1}X_{2k+1}\right)Y_{2j+1},\, \, j= 0,\ldots,L-2,\\
\gamma_{2L-2}&= Z_{2L-1}, \,\, \gamma_{2L-1}=-Y_{2L-1}.
\end{align}
\end{subequations}
The Majorana fermions are constructed using Pauli strings that start from site $2L-1$ which hosts the mixed coupling term. Thus, only the Majorana fermions at site $2L-1$ are defined without a Pauli string. The Majorana fermions satisfy the usual anti-commutation relations $\{\gamma_{2j},\gamma_{2k}\}=\{\gamma_{2j+1},\gamma_{2k+1}\}=2\delta_{kl}$ and $\{\gamma_{2j},\gamma_{2k+1}\}=0$. With these definitions, all the Majorana fermions can be shown to commute with the symmetry operator \eqref{Omdef} except for $\gamma_{2L-2}$ which anti-commutes with it,
\begin{subequations}
\begin{align}
    &\Omega \gamma_{2L-2} = - \gamma_{2L-2} \Omega,\\
    &\Omega \gamma_{2j} = \gamma_{2j}\Omega,\, j=0,\ldots,L-2, \\
    &\Omega \gamma_{2j+1} = \gamma_{2j+1}\Omega,\, j=0,\ldots,L-1.
\end{align}
\end{subequations}
From \eqref{JW} it follows that

\begin{align}
& \gamma_{0} = \left[X_{2L-1}\right]Z_1,\,\, \gamma_1 = -\left[X_{2L-1}\right]Y_1,\nonumber\\
&\gamma_{2} = \left[X_{2L-1}X_1\right]Z_3,\,\, \gamma_3 = -\left[X_{2L-1}X_1\right]Y_3,\nonumber\\
&\gamma_{4} = \left[X_{2L-1}X_1 X_3\right]Z_5,\,\, \gamma_5 = -\left[X_{2L-1}X_1 X_3\right]Y_5,\nonumber\\
&\vdots\nonumber \\
&\gamma_{2L-4} = \left[X_{2L-1}X_1 X_3\ldots X_{2L-5}\right]Z_{2L-3},\nonumber\\
&\gamma_{2L-3} = -\left[X_{2L-1}X_1 X_3\ldots X_{2L-5}\right]Y_{2L-3}.
\end{align}
Substituting the above in \eqref{Omdef}, one obtains $\Omega=\biggl[(-i) \gamma_0 \gamma_1 (-i) \gamma_2 \gamma_3 \ldots (-i) \gamma_{2L-4}\gamma_{2L-3}\biggr]\gamma_{2L-1}$. Thus $\Omega$ contains an odd number of Majorana fermions, i.e, it contains all the Majoranas except $\gamma_{2L-2}$. The Floquet unitary \eqref{Tsig} can now be written as 
\begin{align}\label{TsigM}
&T_{\sigma, 2L-1} = e^{-\sum_{j=0}^{L-2}u_{2j+1}\gamma_{2j}\gamma_{2j+1}}e^{-u_{2L-2}\Omega \gamma_{2L-3}\gamma_{2L-1}}\nonumber\\
&\times e^{-\sum_{j=0}^{L-2}u_{2j}\gamma_{2j-1}\gamma_{2j}},
\end{align}
where the Majorana fermion $\gamma_{2L-2}$ is noticeably absent in the Floquet unitary.
Despite this, $\gamma_{2L-2}$ does not  commute with the Floquet unitary because of the presence of the symmetry operator $\Omega$ attached to the mixed coupling term, and because $\Omega$ anticommutes with $\gamma_{2L-2}$. Since all the remaining Majoranas commute with $\Omega$, we can study the dynamics in a given eigensector of $\Omega$, and in this case $T_{\sigma, 2L-1}$ is entirely bilinear in Majorana fermions, and hence integrable. 

In addition, the Floquet unitary has to have a determinant of $1$, and given that the Floquet unitary contains an odd number of fermions, this implies it has to have  a single unity eigenvalue, namely a single Majorana zero mode. This mode commutes with the time evolution operator, while its ``partner" $\gamma_{2L-2}$ (the other Majorana operator which does not explicitly appear in $T_{\sigma, 2L-1}$), does not. Depending on the coupling parameters, this Majorana zero mode could be delocalized, for example at the self-dual point where $u_{2j+1}=u_{2j}=\text{constant}$, or localized. 
It was shown that for uniform couplings $u_{2j+1}= gT/2, u_{2j}= JT/2$, the Majorana zero mode is localized at the duality twist, with a localization length of \cite{Tan22}
\begin{align}
    \xi \sim \biggl[\ln\bigg\{\frac{\tan(\text{max}(g,J))}{\tan(\text{min}(g,J))}\biggr\}\biggr]^{-1}.
\end{align}

For the sake of clarity, let us construct the Majorana zero mode for two limiting cases. The first case is when $u_{2j}=0$. In this case, the system only has transverse fields, and there are many conserved quantities corresponding to
all the $X_{2j+1}$, as well as $Y_{2L-1}$ and $Z_{2L-1}$ due to the absent transverse field at site $2L-1$. In this case, the spectrum of the many-particle Floquet operator has degeneracies because
we have the following operator algebra: $\left[T_{\sigma,2L-1},\Omega\right]=0, \left[T_{\sigma, 2L-1}, Z_{2L-1}\right]=0$ and $\{\Omega, Z_{2L-1}\}=0$. Thus the operator $Z_{2L-1}$ when acting on the 
$\Omega=1$ eigenstate with a given quasi-energy, changes it to the $\Omega=-1$ eigenstate with the same quasi-energy. When an infinitesimal Ising interaction is switched on, only one zero mode survives, which has the largest overlap with $Y_{2L-1}$ (i.e. $\gamma_{2L-1}$). 
This zero mode commutes rather than anticommutes with $\Omega$ and therefore does
not imply a degeneracy of the Floquet spectrum. 
In general, in contrast to open boundary conditions \cite{FendleyXYZ,Sen13,Yates19}, the degeneracy of the many-particle Floquet spectrum with a duality twist does not survive for generic couplings \cite{Tan22}. However in the high frequency limit,
a Kramers' degeneracy does emerge \cite{Fendley16}.

The second limiting case is when all the transverse fields are zero $u_{2j+1}=0$, while all the exchange interactions are uniform and equal, $u_{2j}=u$. For this case the Majorana zero mode is \cite{Tan22}
\begin{align}
&\Psi = \frac{1}{\sqrt{1 +\sec^2{u}}}\biggl[\Omega \gamma_{2L-3} + 
\tan{u}\, \gamma_{2L-1} + \gamma_0\biggr]\nonumber\\
& = \frac{1}{\sqrt{1 +\sec^2{u}}}\biggl[Z_{2L-3} Z_{2L-1} - 
\tan{u}Y_{2L-1} + X_{2L-1}Z_1\biggr].\label{Mg0}
\end{align}
Deviations from this fine-tuned point do not change the fact that the single site operator $Y_{2L-1}=-\gamma_{2L-1}$ has an overlap with the
Majorana zero mode. 
In later sections, we will use the auto-correlation function of $Y_{2L-1}$ to numerically detect the Majorana zero mode. 

\subsection{Spatial translation of the duality twist}

We now show that a unitary transformation can move the duality defect (or twist) from one link to another, and in doing so,  the role of $u_{\rm even}$ and $u_{\rm odd}$ are interchanged, introducing a ``domain wall'' in the space of coupling parameters.

In what follows, starting with a duality twist located in between the sites $2L-3, 2L-1$ (corresponding to the unitary \eqref{Tsig}), we
will perform a unitary transformation which moves the duality twist to the link between sites $2L-5,2L-3$. In particular, before the action of the unitary transformation, there was no magnetic field at site $2L-1$ 
and there were exchange terms of the kind  $Z_{2L-3}X_{2L-1}, Z_{2L-5}Z_{2L-3}$. The unitary transformation will remove the magnetic field on site $2L-3$, while reintroducing a magnetic field on site $2L-1$. In addition, the unitary transformation will change the exchange terms to $Z_{2L-5}X_{2L-3}$ and $Z_{2L-3} Z_{2L-1}$. The couplings on all other sites will remain unchanged. 

In order to achieve this, let us define the controlled-Z  (CZ) gate
\begin{align}
CZ_{1,2} = |0\rangle\langle 0|_1\otimes 1_{2} 
+  |1\rangle\langle 1|_1\otimes Z_{2}.
\end{align}
We will now show that the following unitary transformation \cite{Fendley16}
\begin{align}
U_{D_{\sigma},2L-1} =CZ_{2L-3,2L-1}H_{2L-3},
\end{align}
will implement the required transformation where $H_j = (X_j + Z_j)/\sqrt{2}$ is the Hadamard gate on site $j$.

First note that the action of the CZ gate on the unitary in \eqref{Tsig} is
\begin{align}
&CZ_{2L-3,2L-1}^{\dagger} \biggl[T_{\sigma, 2L-1} \biggr] CZ_{2L-3,2L-1} \nonumber\\
&= \left(\prod_{j=0}^{L-3}W_{2j+1}^{X}(u_{2j+1})
 \right)\nonumber\\
 &\times e^{-i u_{2L-3}X_{2L-3}Z_{2L-1}} e^{-iu_{2L-2}X_{2L-1}}   \left(\prod_{j=0}^{L-2}
 W_{2j}^{ZZ}(u_{2j})\right). 
\end{align}
In particular, the
following transformations 
$X_{2L-3} \rightarrow X_{2L-3}Z_{2L-1}$ and $Z_{2L-3}X_{2L-1} \rightarrow X_{2L-1}$, have been implemented. 
At the next step the Hadamard gate transforms $X_{2L-3} (Z_{2L-3})$
to $Z_{2L-3}(X_{2L-3})$ giving the transformed unitary
\begin{align}
&T_{\sigma,2L-1}'=U_{D_{\sigma},2L-1}^\dagger T_{\sigma,2L-1} U_{D_{\sigma},2L-1}\nonumber\\
&=\biggl[\left(\prod_{j=0}^{L-3}W_{2j+1}^{X}(u_{2j+1})
 \right)\nonumber\\
 &\times e^{-i u_{2L-3}Z_{2L-3}Z_{2L-1}}  e^{-iu_{2L-4}Z_{2L-5}X_{2L-3}}e^{-iu_{2L-2}X_{2L-1}}\nonumber\\
&\times  \left(\prod_{j=0}^{L-3}
 W_{2j}^{ZZ}(u_{2j})\right)\biggr].
\end{align}
This realizes what we set out to do. In particular, after the unitary transformation there is no magnetic field on site $2L-3$, a magnetic field on all other sites and a $Z_{2L-5}X_{2L-3}$ twist. Note that the translation of the twist has locally interchanged an Ising coupling with a transverse field coupling. This is reflected in $u_{2L-2}$ now representing a transverse field coupling, while 
$u_{2L-3}$ becomes an Ising coupling.  
Now let us explore what happens to the symmetry $\Omega$ in
\eqref{Omdef}. It is straightforward to see that
\begin{align}
   & U_{D_{\sigma},2L-1}^{\dagger} \Omega U_{D_\sigma,2L-1} = i Z_{2L-3}D_{\psi}.
\end{align}

To move the twist further to the link between sites $2L-7, 2L-5$, we need to apply
\begin{align}
U_{D_{\sigma},2L-3} =CZ_{2L-5,2L-3}H_{2L-5}.
\end{align}
The above CZ gate will convert $Z_{2L-5}X_{2L-3} \rightarrow X_{2L-3}$  and
$X_{2L-5} \rightarrow X_{2L-5}Z_{2L-3}$. After application of the
Hadamard gate on site $2L-5$ one obtains the new unitary 
\begin{align}
&    T''_{\sigma, 2L-1} =U^{\dagger}_{D_{\sigma},2L-3} T'_{\sigma, 2L-1} U_{D_{\sigma},2L-3}\nonumber\\
    &=\biggl[\left(\prod_{j=0}^{L-4}W_{2j+1}^{X}(u_{2j+1})
 \right) \nonumber\\
 &\times e^{-i u_{2L-3}Z_{2L-3}Z_{2L-1}} e^{-i u_{2L-5}Z_{2L-5}Z_{2L-3}}\nonumber\\
 &\times e^{-i u_{2L-6}Z_{2L-7}X_{2L-5}} e^{-iu_{2L-2}X_{2L-1}} e^{-iu_{2L-4}X_{2L-3}} \nonumber\\
&\times  \left(\prod_{j=0}^{L-4}
 W_{2j}^{ZZ}(u_{2j})\right)  \biggr].
\end{align}
We now see that the $u_{\rm even}$ and
$u_{\rm odd}$ have exchanged roles for two additional spins located on sites $2L-5, 2L-3$. Moreover there is no transverse field on site $2L-5$.

Quite generally, the unitary transformation
\begin{align}
 U_{D_\sigma,2j+1}:=CZ_{2j-1,2j+1}H_{2j-1}, 
\end{align}
moves the duality link from between
sites $2j-1, 2j+1$ to between sites
$2j-3,2j-1$. After this transition, what was originally a transverse-field coupling $u_{2j-1}$ becomes the Ising coupling between sites $2j-1$ and $2j+1$, while the original mixed coupling term $u_{2j}$ becomes the
transverse field coupling for the spin at site $2j+1$. Denoting the term ``domain wall'' for the point where the Ising  and transverse fields exchange roles, 
the very first translation introduces a domain wall, and each subsequent translation increases the size of one ``phase'' relative to the other. This is highlighted pictorially in Fig.~\ref{shift}.

It should be noted that unitary transformations that move the duality twist have been constructed for the quantum Hamiltonian at the critical point \cite{Vidal16}, as well as for the classical 2D Ising system at and away from the critical point \cite{Fendley16}.
In this section we have extended previous work by generalizing the unitary transformation to Floquet systems. 

As we apply the unitary transformation to move the duality twist, the Majorana zero mode remains localized at the domain wall. To show this, we take as an example the exactly solvable case where all the transverse fields are zero and all the exchange interactions are equal. The zero mode is given in (\ref{Mg0}). After we move the duality twist from between sites $2L-3$ and $2L-1$ to between sites $2k-3$ and $2k-1$, the zero mode (expressed in spin degrees of freedom) gets transformed as
\begin{equation}
\begin{split}
 \Psi'&=U^\dagger_{D_\sigma,2k+1}\dots  U^\dagger_{D_\sigma,2L-1}\Psi U_{D_\sigma,2L-1}\dots U_{D_\sigma,2k+1}\\
 &=\frac{1}{\sqrt{1+\sec^2{u}}}\left[ -\tan{u} \prod_{j\geq k}^{L-1}X_{2j-1} Y_{2L-1} \right.\\
 &\left.+\prod_{j\geq k}^{L-1}X_{2j-1} Z_{2L-1}
 +\prod_{j\geq k}^{L}X_{2j-1} Z_1\right].
 \end{split}
\end{equation}  In particular, the zero mode consists of string operators spanning between the duality twist and the domain wall at site $2L-1$. Our convention for the Jordan-Wigner transformation is such that Majorana fermions correspond to string operators attached to the duality twist. In terms of these Majorana fermions, the zero mode becomes
\begin{equation}
\Psi= \frac{1}{\sqrt{1+\sec^2{u}}}\left[ \gamma'_{2L-2} +\tan{u} \gamma'_{2L-1}+\gamma'_0 \right],
\end{equation}
where we denote the Majorana fermions as $\gamma'$ to distinguish from the case where the duality twist is fixed to be between sites $2L-3$ and $2L-1$. The Majorana zero mode is indeed localized around the domain wall at site $2L-1$ and it remains there when one moves the location of the duality twist using the procedure described in Fig.~\ref{shift}.

For a different Floquet unitary from the one above, it was shown that when a domain wall exists, the Majorana mode is located at the domain wall \cite{Tan22}.
In the ground state
of a  chain described by a time-independent Hamiltonian (equivalent to the high frequency limit of the Floquet chain), a domain wall configuration of coupling constants would separate a paramagnetic phase from a ferromagnetic phase,  with the Majorana zero
mode residing at the boundary of the two phases, i.e at the domain wall \cite{Fendley16}.
In particular, the duality twist allows one to introduce an odd number of domain walls, unlike periodic boundary conditions which always require even number of domain walls. In the Floquet context, while there is no phase, a Majorana mode nevertheless exists, localized at the domain wall across which $u_{\rm even}, u_{\rm odd}$ interchange roles. 

\begin{figure}
\includegraphics[width = 0.45\textwidth]{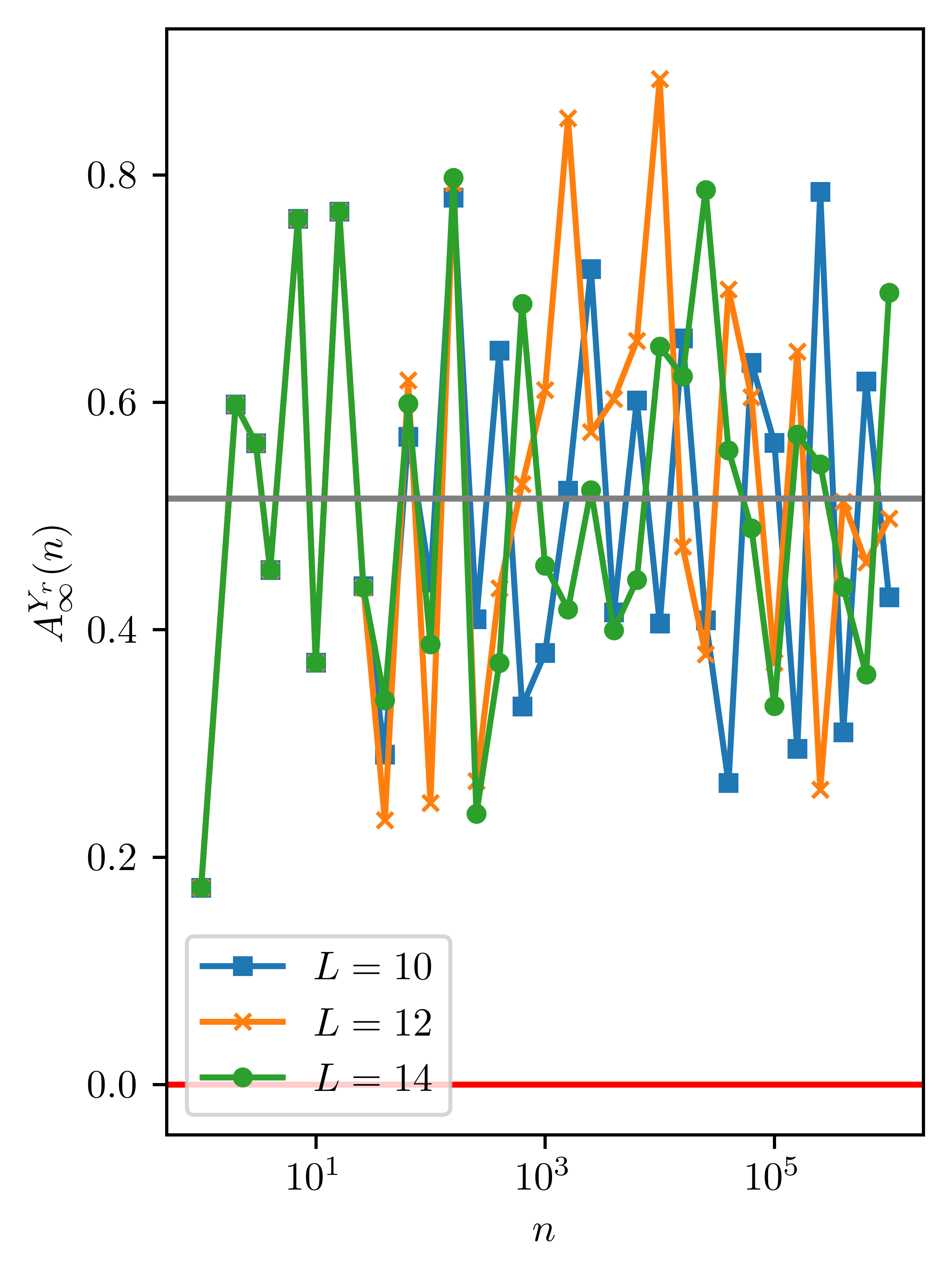}
\caption{\label{Jx0Y} Autocorrelation function \eqref{AYr}
where $Y_r$ has an overlap with the Majorana zero mode, and for the integrable case $J_x=0$. The other parameters are
$gT/2=0.3$, $JT/2=1.0$. Grey lines
are the numerically constructed plateau heights $P$ \eqref{Pdef} related to the
conserved quantity $Q$ \eqref{Qdef}. When $J_x=0$, $P$ shows little dependence on $L$.}
\end{figure}

\begin{figure}
\includegraphics[width = 0.45\textwidth]{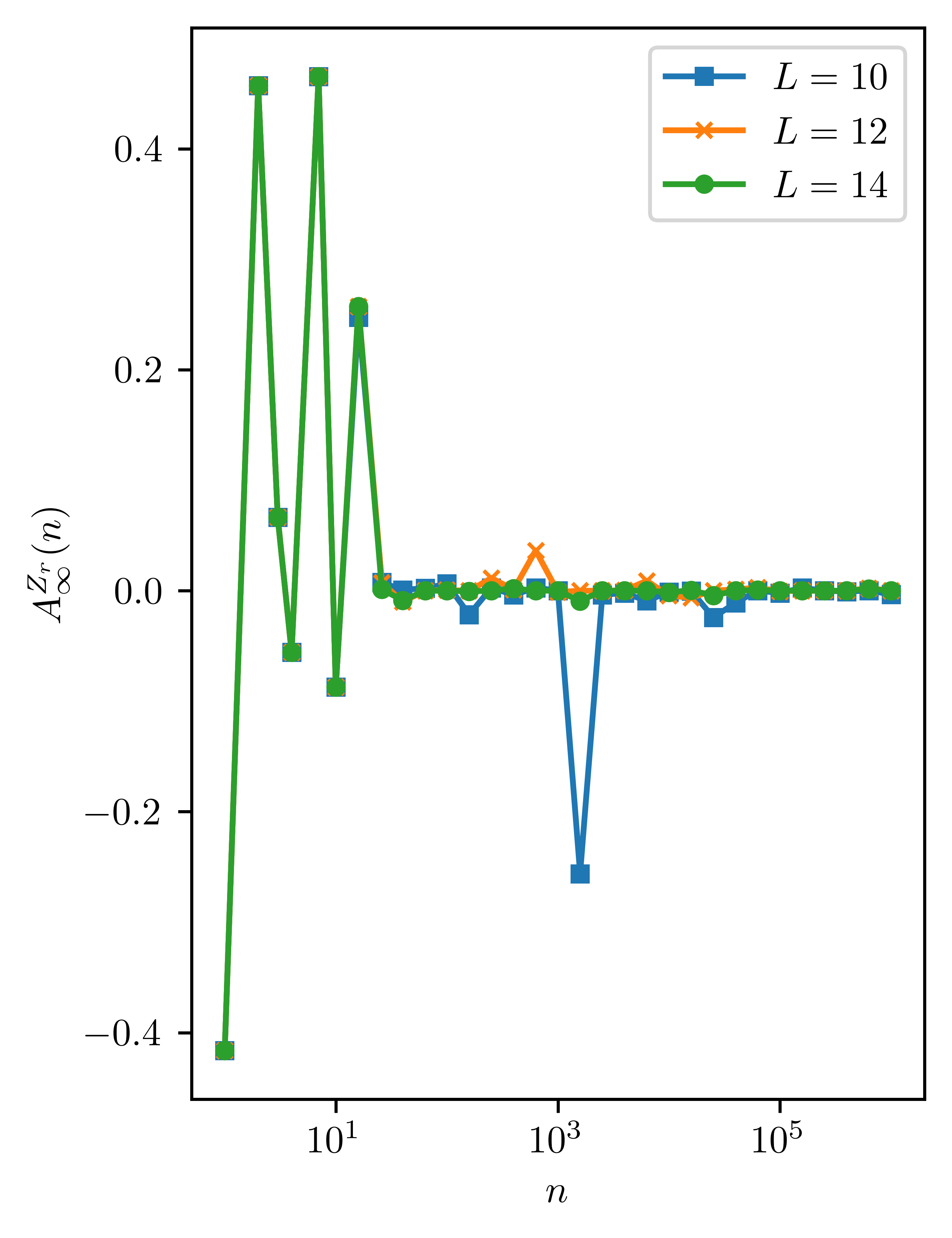}
\caption{\label{Jx0Z}  Autocorrelation function \eqref{AZr}
where $Z_r=\gamma_{2L-2}$. This is the Majorana operator that is left out of the integrable Floquet unitary with $J_x=0$. The other parameters are
$gT/2=0.3$, $JT/2=1.0$.}
\end{figure}

\subsection{Dynamics of $\gamma_{2L-2}$}
The peculiarity of the duality twisted Floquet unitary in \eqref{TsigM} is that while there are $2L$ Majorana fermions,
only an odd number ($2L-1$) of them appear explicitly, with the Majorana operator $\gamma_{2L-2}$ absent from $T_{\sigma, 2L-1}$. In addition, since all the Majorana fermions except $\gamma_{2L-2}$ commute with the symmetry $\Omega$, they transform in a standard manner in that they evolve into linear superpositions of single Majorana fermions.
Denoting $\vec{\gamma} = \left[\gamma_0,\gamma_1, \gamma_2, \ldots, \gamma_{2L-3}, \gamma_{2L-1}\right]$ as a $2L-1$
component vector, this vector evolves as 
\begin{equation}
    T_{\sigma, 2L-1}^\dagger \vec{\gamma} T_{\sigma,2L-1} = M_\Omega \vec{\gamma},
\end{equation}
where $M_{\Omega} $ is a $(2L-1) \times (2L-1)$ orthogonal matrix which depends on $\Omega$. Thus as long as one is interested in dynamics involving the above
$2L-1$ Majorana fermions, the dynamics is integrable in that it scales linearly with system size $L$.

In contrast, the Majorana fermion $\gamma_{2L-2}$, due to its anticommutation with $\Omega$, has a rather complicated time-evolution given by \cite{Tan22}
\begin{align}
&T_{\sigma,2L-1}^\dagger \gamma_{2L-2} T_{\sigma,2L-1} = \cos{2u_{2L-2}}\gamma_{2L-2}\nonumber\\
&+\Omega \sin{2u_{2L-2}} \cos{2u_0}\gamma_{2L-2}\gamma_{2L-3}\gamma_{2L-1} \nonumber \\
    &-\Omega \sin{2u_{2L-2}}\sin{2u_0} \gamma_{2L-2}\gamma_{2L-3}\gamma_0.
\end{align}

Let us use the notation
\begin{align}
\left[T_{\sigma, 2L-1}^\dagger\right]^n \gamma_{i}(0) \left[T_{\sigma, 2L-1}\right]^n = \gamma_i(n),
\end{align}
and define the operator
\begin{align}
O &= \biggl[ \cos{2u_{2L-2}}-\Omega \sin{2u_{2L-2}} \cos{2u_0}\gamma_{2L-3}\gamma_{2L-1} \nonumber\\
&    +\Omega \sin{2u_{2L-2}}\sin{2u_0} \gamma_{2L-3}\gamma_0\biggr], \label{Odef}
\end{align}
then,
\begin{align}
T_{\sigma, 2L-1}^\dagger \gamma_{2L-2}(0) T_{\sigma, 2L-1} =\gamma_{2L-2}(0) O(0).
\end{align}
Thus, after two Floquet cycles,
\begin{align}
\biggl(T_{\sigma, 2L-1}^\dagger\biggr)^2 \gamma_{2L-2} \biggl(T_{\sigma, 2L-1}\biggr)^2 = \gamma_{2L-2}(0) O(0) O(1),
\end{align}
while after $n$ Floquet periods
\begin{align}
&\biggl(T_{\sigma, 2L-1}^\dagger\biggr)^n \gamma_{2L-2} \biggl(T_{\sigma, 2L-1}\biggr)^n \nonumber\\
&= \gamma_{2L-2}(0) O(0) O(1) O(2) \ldots O(n-1).\label{eq:stringCorr}
\end{align}
Thus under Floquet time-evolution, the $\gamma_{2L-2}$ operator evolves into longer and longer strings of Majorana fermions. The dynamics is therefore non-trivial  if it involves $\gamma_{2L-2}$. We may employ exact diagonalization (ED) to study how this Majorana evolves, and we will demonstrate its explicit time-evolution in the subsequent sections. However, it is interesting to note that one gets a type of X-ray edge problem when calculating the time-evolution of 
$\gamma_{2L-2}$ because $\gamma_{2L-2}$ takes $\Omega \rightarrow -\Omega$. In particular, we have
\begin{align}\label{xray}
&\biggl(T_{\sigma, 2L-1}^\dagger\biggr)^n \gamma_{2L-2} \biggl(T_{\sigma, 2L-1}\biggr)^n\nonumber\\
&=\gamma_{2L-2} 
\biggl(T^{\dagger}_{\sigma, 2L-1}(-\Omega)\biggr)^n \biggl(T_{\sigma,2L-1}(\Omega)\biggr)^n.
\end{align}
In the above equation, the notation $T_{\sigma, 2L-1}(\pm \Omega)$ keeps track of which $\Omega$ sector the unitary corresponds to.

\section{Non-integrable Duality Twisted Unitary} \label{sec:NInt}

In the remaining sections  we will study the system by performing ED on chains of  
length $L=4,6,8,10,12,14$. Since the sites are labelled as $0,1,2,\ldots 2L-1$ with 
the spins on the odd sites, the number of
spins equals $L$. In addition, we will place the 
duality twist at the center of the chain, i.e at $r = L-1 \in $ odd. 
Thus
there is no transverse field on the spin at site $r$,
and the exchange couplings between sites $r-2,r$ is of the mixed coupling type $Z_{r-2}X_r$.  To keep the discussion simple, there will be no
domain walls in the coupling configuration space.
The Jordan-Wigner transformation starts from site $r$ so that
\begin{align}
\gamma_{r-1}&= Z_{r}, \,\, \gamma_{r}=-Y_{r}.
\end{align}
Thus it is now $Y_r$ that has an overlap with the zero mode, and it is $Z_r$
that equals the Majorana fermion that does not explicitly enter the Floquet unitary. 
As before, periodic boundary conditions where $O_{2L+n} = O_{2L}$, is implied. 

We choose all the
non-zero transverse field couplings to be uniform and equal to $g T/2$, and the $ZZ$ and $ZX$ Ising couplings are also chosen to be uniform and equal to $J T/2$. 
The discrete $\mathbb{Z}_2$ symmetry operator with the above labeling of sites is
\begin{align}
    \Omega = - X_1 X_3 \ldots X_{r-2}Y_r X_{r+2} \ldots X_{2L-1}
    . \label{Omdef2}
\end{align}

We also break integrability by introducing a nearest-neighbour exchange interaction of the type $X_{2j-1} X_{2j+1}, j \in$ integer, which after a Jordan-Wigner transformation corresponds to four fermion interactions. We
consider two kinds of unitaries. One is $T_{\sigma, S}$ where the unitary still commutes with $\Omega$. We achieve this by 
introducing $XX$ interactions on all odd nearest-neighbor sites except between sites $r-2,r$ and $r, r+2$. The second unitary we consider is
$T_{\sigma, A}$ where the $X X$ type exchange interaction exists between all odd nearest-neighbor sites, and therefore $\Omega$ is no longer a symmetry.
Concretely, the unitary that preserves the $Z_2$ symmetry is
\begin{align}
    T_{\sigma,S}&= e^{-i (J_x T/2) \sum_{j\neq (r-1)/2,(r+1)/2,j=1}^{L}X_{2j-1}X_{2j+1}} \nonumber\\ 
    &\times e^{-i (g T/2) \sum_{j\neq (r-1)/2, j=1}^{L}X_{2j+1}} \nonumber\\
    &\times e^{-i (J T/2) Z_{r-2}X_{r}}e^{-i (JT/2) \sum_{j\neq (r-1)/2, j=1}^{L}Z_{2j-1}Z_{2j+1}}
    \label{TinS},
\end{align}
while the one with no $Z_2$ symmetry is
\begin{align}
    T_{\sigma,A}&= e^{-i (J_x T/2) \sum_{j=1}^{L}X_{2j-1}X_{2j+1}} \nonumber\\ 
    &\times e^{-i (g T/2) \sum_{j\neq (r-1)/2, j=1}^{L}X_{2j+1}} \nonumber\\
    &\times e^{-i (J T/2) Z_{r-2}X_{r}}e^{-i (JT/2) \sum_{j\neq (r-1)/2, j=1}^{L}Z_{2j-1}Z_{2j+1}}
    \label{TinA},
\end{align}
Above, site $2L+1$ is identified with site $1$ and so
an exchange term of the kind $Z_{2L-1} Z_{2L+1}$ is the same as $Z_{2L-1} Z_1$, and the exchange term $X_{2L-1} X_{2L+1}$ is the same as $X_{2L-1} X_1$. $J_x$ parametrizes the strength of the integrability breaking interactions.

Let us define an infinite-temperature autocorrelation function of the Pauli spins $Y_r, Z_r$ at site $r$ at stroboscopic time $n$ as 
\begin{align}
    A_{\infty}^{Y_r}(n) &= \frac{1}{2^L}
    {\rm Tr} \biggl[Y_r(n) Y_r(0)\biggr], \label{AYr}\\
A_{\infty}^{Z_r}(n) &= \frac{1}{2^L}
    {\rm Tr} \biggl[Z_r(n) Z_r(0)\biggr] \label{AZr}.
\end{align}
It is $Y_r$ that has an overlap with the Majorana zero mode in the integrable limit of $J_x=0$, and therefore stability of the zero mode will be explored by the study of $A_{\infty}^{Y_r}$.
In addition, the left-out Majorana in the integrable limit is identical to $Z_r$, and therefore its unusual dynamics, even in the integrable case, will be explored through the study of $A_{\infty}^{Z_r}$.

We also find it helpful to construct a numerical conserved quantity with an overlap with
$Y_r$, 
\begin{align}
Q = \sum_{\beta} \langle \beta|Y_r |\beta\rangle  |\beta\rangle \langle \beta|,\label{Qdef}
\end{align}
where $|\beta\rangle$ denote the exact eigenstates of $T_{\sigma, S/A}$. Then we expect 
that if such a conserved quantity exists, the plateau
height of the autocorrelation function of $Y_r$ will be given by
\begin{align}
    P = \langle Q^2 \rangle =\frac{1}{2^L} \sum_{\beta} |\langle \beta|Y_r |\beta\rangle|^2 \label{Pdef}.
\end{align}
with $P = \lim_{n \to \infty} A^{Y_r}_\infty(n)$. We will plot the corresponding $P$ along with the autocorrelation function. 

For the rest of the paper we take
$JT/2=1.0, T=2.0$ while considering different $g$ and $J_x$.
To orient oneself, let us first consider the integrable case of $J_x=0$. 
Fig.~\ref{Jx0Y} shows $A^{Y_r}_{\infty}$
for chains of different lengths and $gT/2=0.3$. The plot shows that this autocorrelation function does not decay with time. The horizontal grey line is the value of $P$ in \eqref{Pdef}. Unlike the
$J_x\neq 0$ cases discussed later, $P$ is not $L$ dependent for the integrable case of $J_x=0$. 
Fig.~\ref{Jx0Z} shows $A^{Z_r}_{\infty}$ for the same parameters, and shows that this decays within time of $O(1)$. This rapid decay can be interpreted as
rapid decoherence due to X-ray edge type processes, see \eqref{xray}, or equivalently due to the non-local string correlation described by \eqref{eq:stringCorr}, building up over time. 

\begin{figure}
\includegraphics[width = 0.45\textwidth]{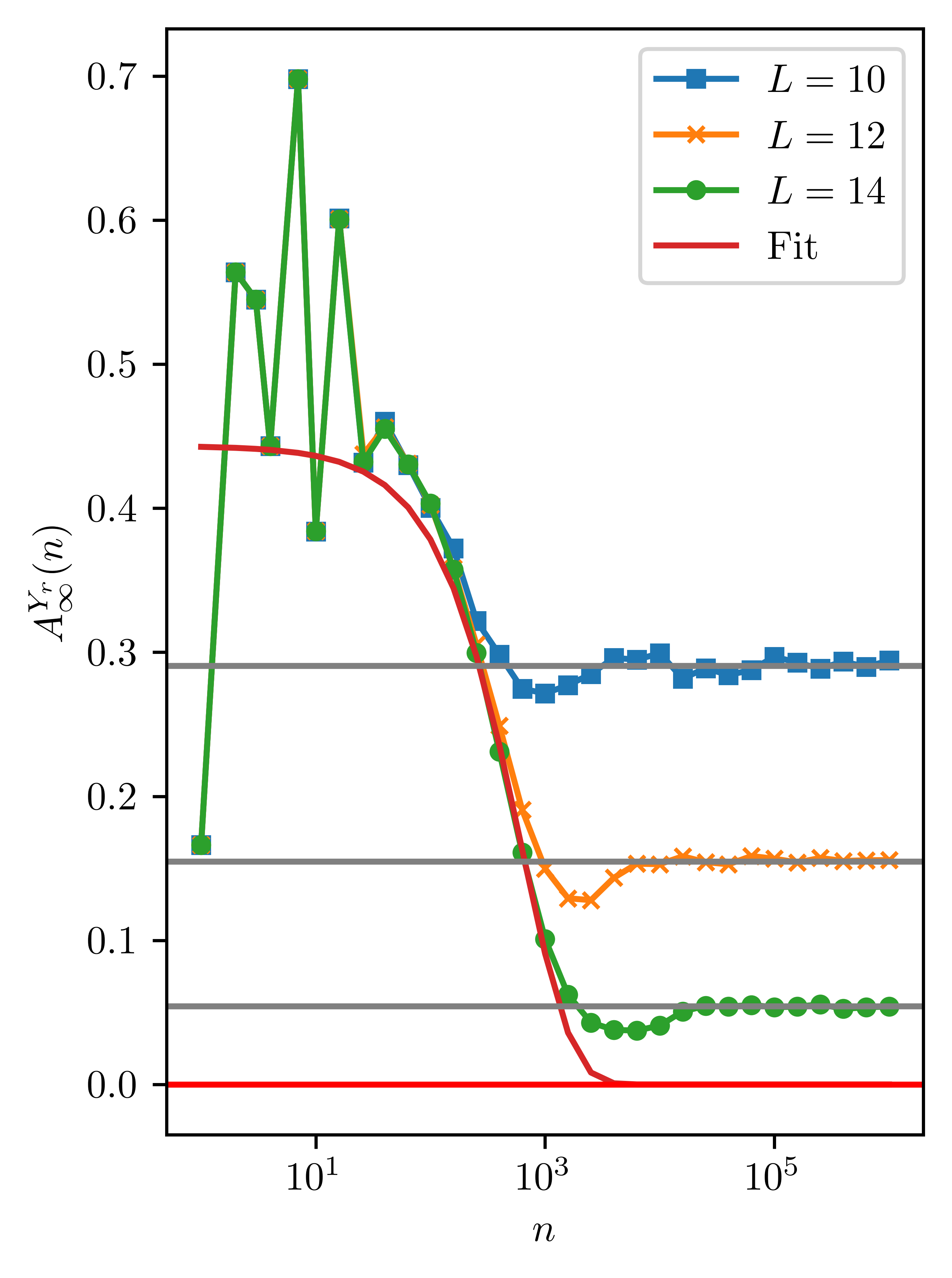}
\caption{\label{Jx0p1}  
The autocorrelation function \eqref{AYr} for the non-integrable case where $gT/2=0.3, JT/2=1.0, T=2.0, J_x=0.1$, and there is no $Z_2$ symmetry. 
Plateaus agree well with the grey lines which 
are the numerically constructed plateau heights $P$, related to the
conserved quantity $Q$. The red line is a fit to an exponential decay $\sim \exp(-\text{Decay-Rate}\, n)$.}
\end{figure}

We now break integrability by including four fermion interactions. Our goal is to explore the fate of the Majorana zero mode. Fig.~\ref{Jx0p1}  presents results for $ JT/2=1.0, T=2.0, g=0.3,J_x=0.1$ for the $Z_2$ symmetry breaking case, and for different system sizes.
After an initial transient, $A_{\infty}^{Y_r}$ settles into a plateau. The
height of the plateau decreases as $L$ increases, and agrees very well with the quantity $P$ as defined in \eqref{Pdef}, which is related to the conserved quantity $Q$. 
While plateau heights in Fig.~\ref{Jx0p1} depend on $L$, the $n$ dependence of the correlation function before the plateau is reached, does not depend on $L$. It fits an exponential very well for the chosen parameters. 
This exponential can be used to extract a decay rate.

\begin{figure}
\includegraphics[width = 0.45\textwidth]{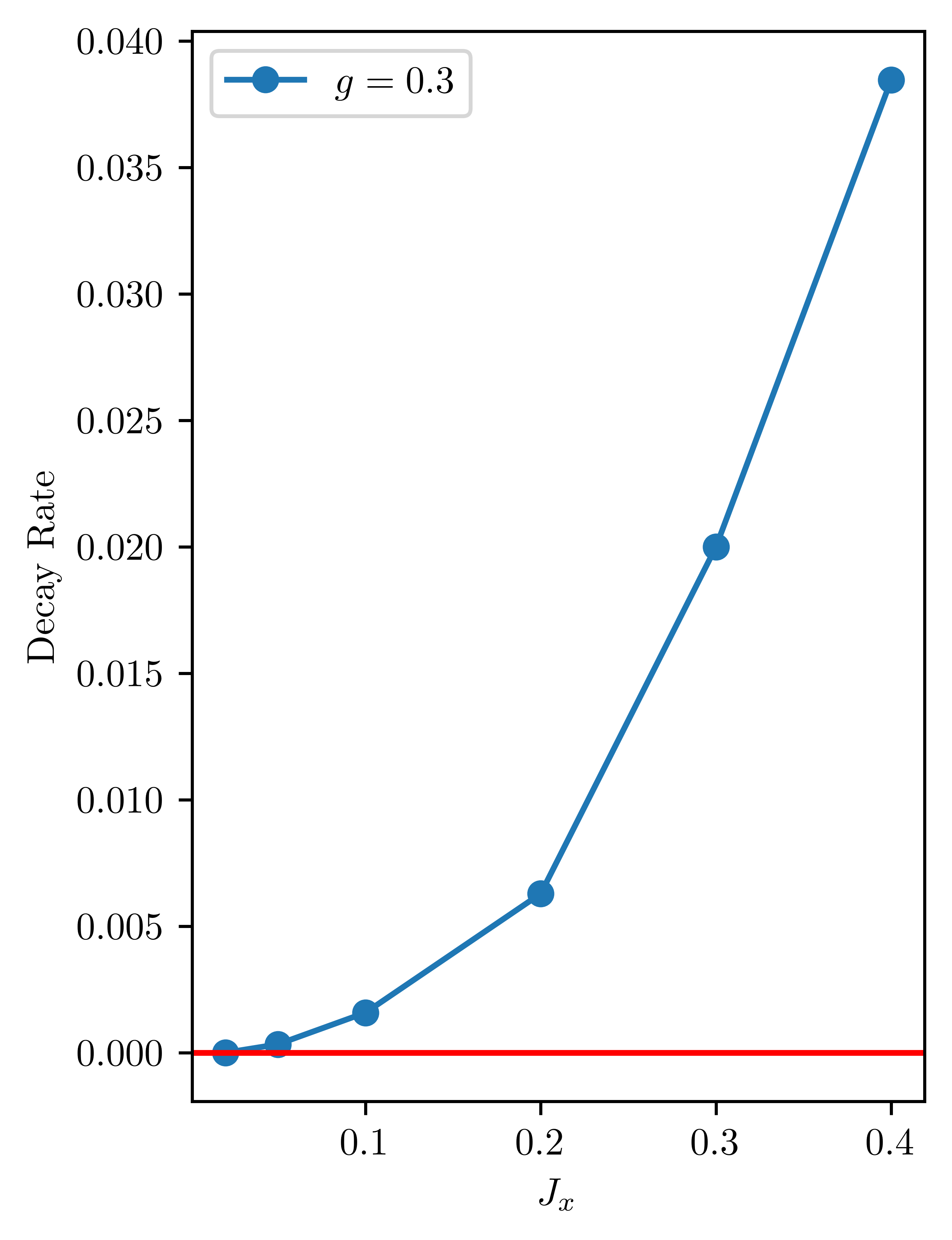}
\caption{\label{decay}  
Decay rates for different $J_x$ for $JT/2=1.0, T=2.0, gT/2=0.3$, and no $Z_2$ symmetry. Decay rates are obtained from fitting the approach to the plateau to an exponential. The decay rates approximately grow with $J_x^2$ for the chosen parameters.}
\end{figure}

The decay-rates for several different $J_x$ and for $JT/2=1.0, T=2.0,g=0.3$ are plotted in Fig.~\ref{decay}. The decay rates indicate a $J_x^2$ dependence, a result also expected from applying Fermi's Golden Rule for the above values of $gT/2, JT/2$ \cite{HCY23}.

\begin{figure}
\includegraphics[width = 0.45\textwidth]{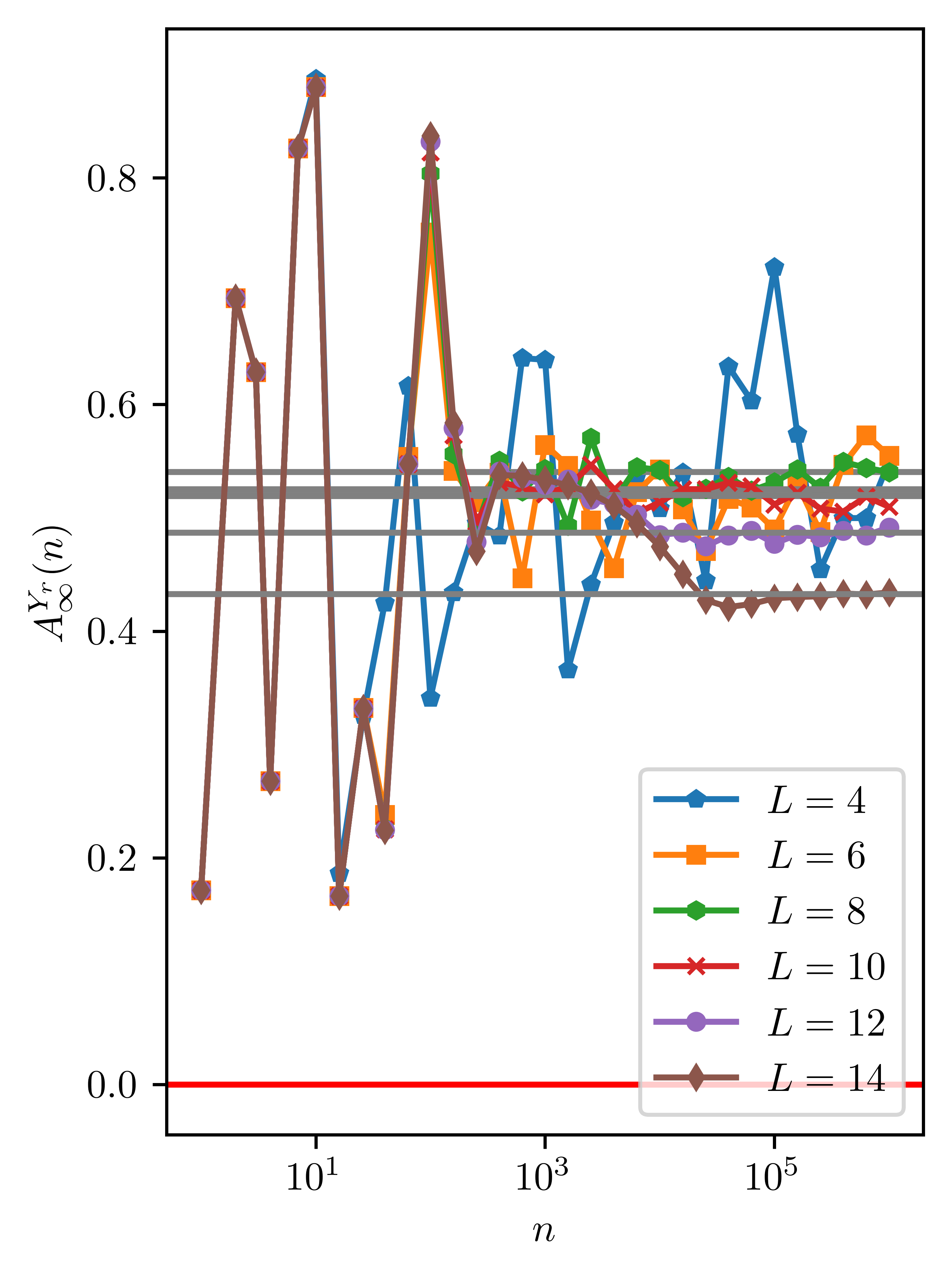}
\caption{\label{ASYM}  
Autocorrelation function for $JT/2=1.0, T=2.0, gT/2=0.1$, $J_x = 0.05$ and no $Z_2$ symmetry. Plateaus agree well with the grey lines which 
are the numerically constructed plateau heights $P$, related to the
conserved quantity $Q$. Smaller transverse fields $g$ and interactions $J_x$ require a larger system size $L$ for seeing a significant drop in the plateau height (compare with Fig.~\ref{Jx0p1}).}
\end{figure}

\begin{figure}
\includegraphics[width = 0.45\textwidth]{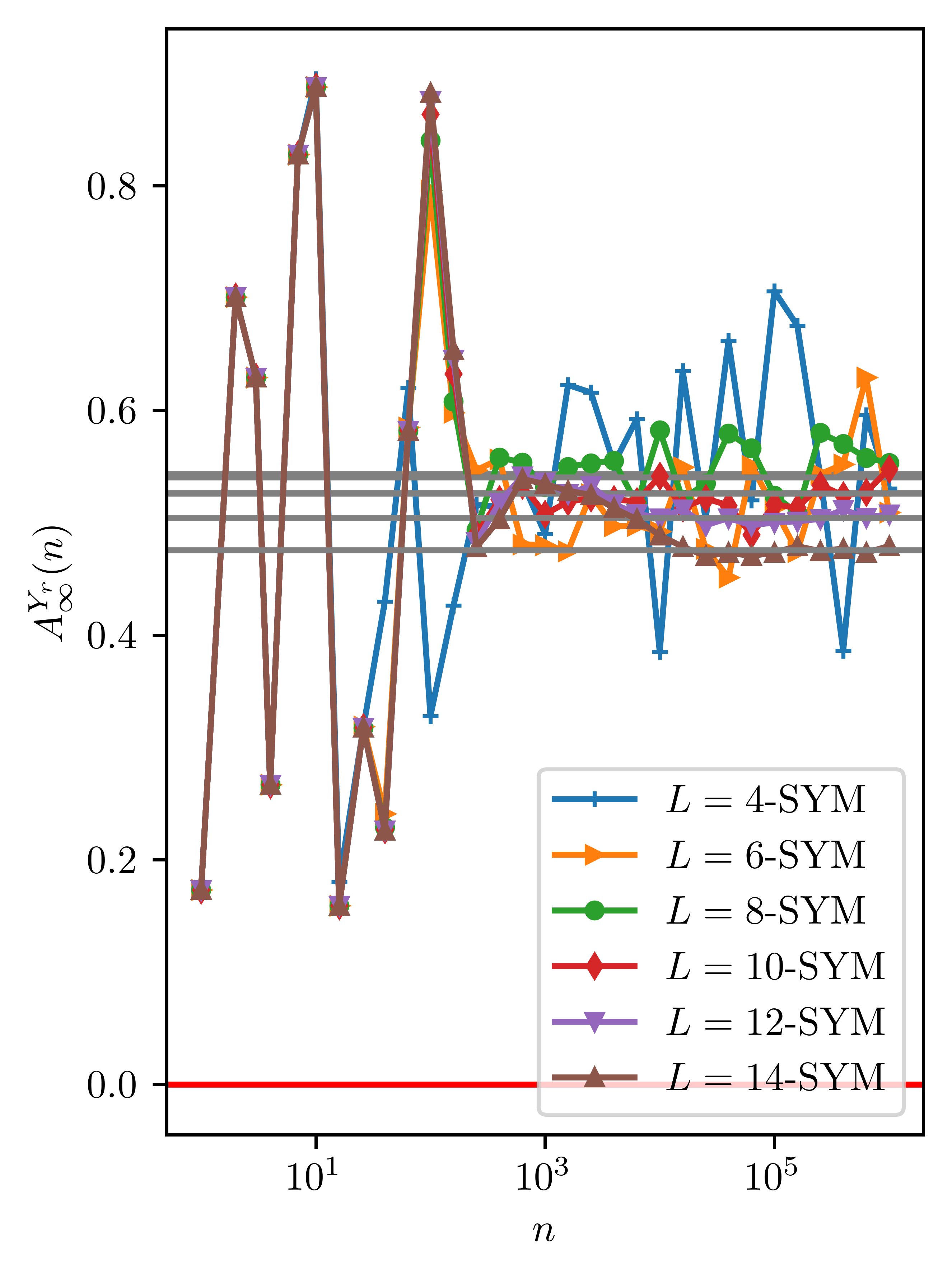}
\caption{\label{SYM}  
Autocorrelation function for $gT/2=0.1$, $JT/2=1.0, T=2.0$, $J_x  = 0.05$ and with $Z_2$ symmetry.
Plateaus agree well with the grey lines which 
are the numerically constructed plateau heights $P$, related to the
conserved quantity $Q$. No qualitative difference is observed between the cases with and without  $Z_2$ symmetry, where $P$ is slightly larger for the former case in comparison to
the latter case (Fig.~\ref{ASYM}), for the same parameters.}
\end{figure}

Quite generally if we make the integrability breaking term smaller and/or the
Majorana mode too localized (for example by reducing the ratio of $\text{min}(g,J)/\text{max}(g,J)$), the system sizes needed to observe a significant decrease in the plateau heights, become too large. This is demonstrated in Figs.~\ref{ASYM} and \ref{SYM} which are both for $gT/2=0.1, JT/2=1.0, T=2.0,J_x=0.05$, but with the former figure for the case with no $Z_2$ symmetry, while the latter for the case with $Z_2$ symmetry. We observe the same qualitative behavior for the two cases.


It is interesting to note that for the integrable case (Fig.~\ref{Jx0Y})
although the plateaus are $L$ independent, the auto-correlation function fluctuates
about the plateau considerably. This is because when the system is integrable, $Y_{2L-1}$ has overlap not just with the Majorana zero mode, but also
with other conserved quantities, such as delocalized quasi-particles. The latter oscillates, causing an oscillatory behavior of the plateau controlled by the single-particle level spacing, proportional to $1/L$. When interactions are included, these delocalized quasi-particles are not conserved, causing the oscillations to dampen rapidly, leaving only the effect of a single conserved quantity, $Q$, and tiny oscillations arising from the exponentially small many-particle level spacing. This is the reason why the oscillations about the plateau value are absent in Fig.~\ref{Jx0p1}.

\section{Plateau height and Fock space delocalization}\label{sec:Pl}

\begin{figure}
\includegraphics[width = 0.45\textwidth]{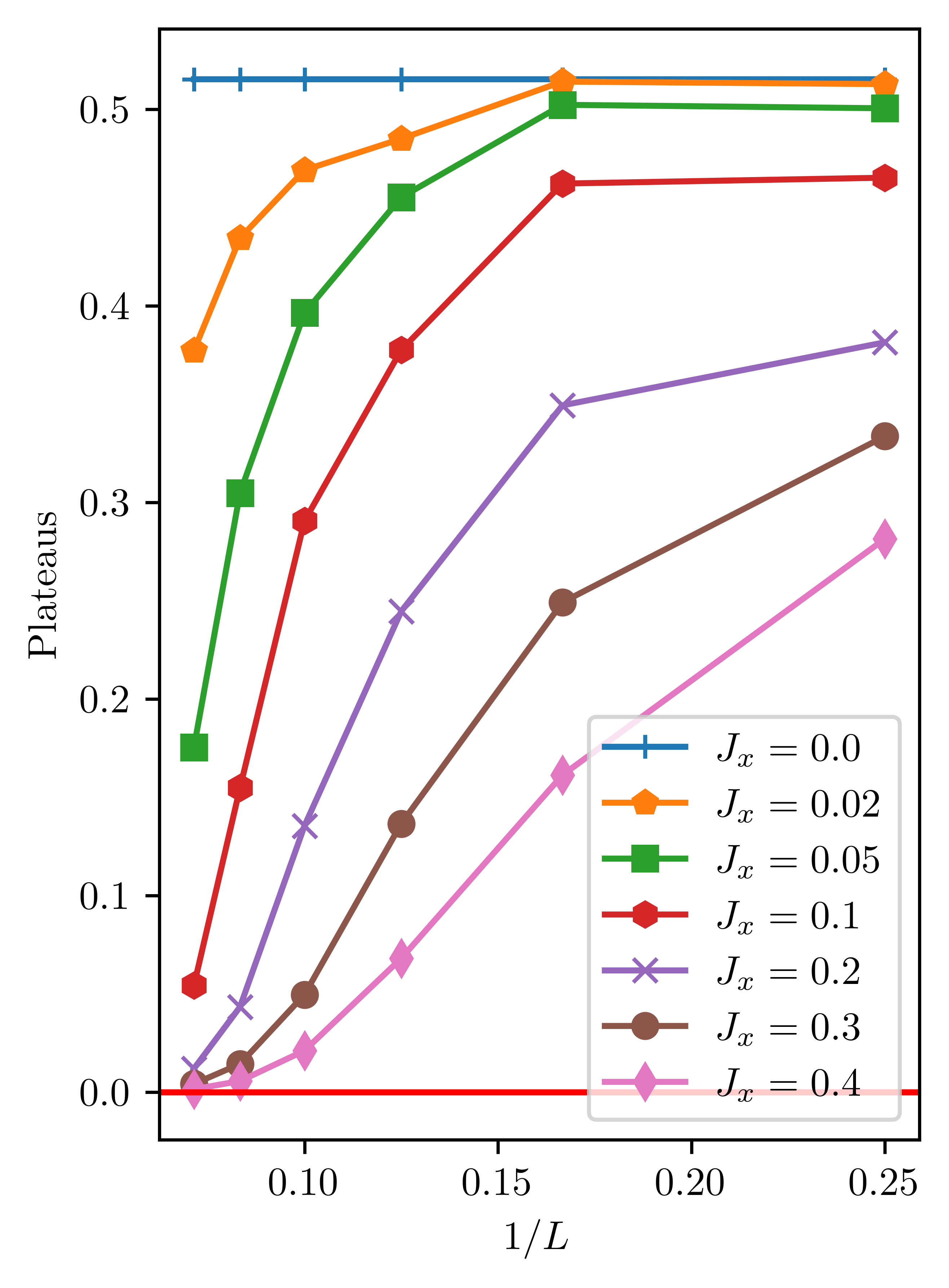}
\caption{\label{P1}  
Plateau heights $P$ \eqref{Pdef} vs $1/L$ for different $J_x$ for the case where there is no $Z_2$ symmetry and $JT/2=1.0, T=2.0, gT/2 = 0.3$.}
\end{figure}

Fig.~\ref{P1} plots the plateau heights $P$ as a function of system size $L$ for the case of no $Z_2$ symmetry, for several different $g, J_x$, with $J T/2=1.0, T=2.0$.  
The plateau height is a finite system size effect because as $L\rightarrow \infty$,
$P\rightarrow 0$. 

\begin{figure}[t]
\includegraphics[width =0.5  \linewidth]{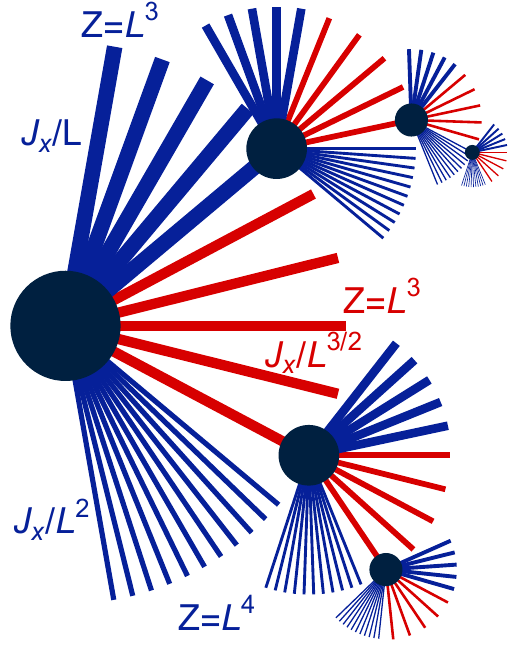}
\caption{\label{fig.flower}  
Interactions $J_x$ induce a hopping from one of the $2^L$ Fock space states (black dots) to the next, if they are connected by
a 4-Majorana process. Thick blue lines: momentum-conserving  scattering processes of extended bulk states with matrix element $M_{\rm ee}\sim J_x/L$. $Z=L^3$ is the number of neighboring states connected by this process. Thin blue lines:  scattering of extended states without momentum conservation, $M^b_{\rm ee}\sim J_x/L^2$. Red lines: interaction terms involving the Majorana zero mode at the boundary/twist, $M_{\rm eM}\sim J_x/L^{3/2}$. For $J_x \ll J/L^2$, Fock states remain localized but upon increasing $J_x$, bulk scattering  leads to a delocalization in Fock space, followed by an activation of edge-Majorana scattering events.
}
\end{figure}
A generic non-integrable interacting many-particle system at finite or, as in our case, infinite temperature is ergodic and equilibrates if it is sufficiently large \cite{Rabson04,Jung2006,Jung2007,Rigol2008,Rigol10,Tavora13,AMreview,Znidaric20,Bulchandani22}.
 The nature of the eigenstates of the time-evolution operator change dramatically as a function of $J_x$. For $J_x=0$, eigenstates are simple many-particle Fock states $|\Omega,\vec n\rangle$, parameterized by $\Omega=\pm 1$ and the occupation numbers $n_i$, $i=1,\dots,L-1$  of the fermionic modes of the system. Using this notation, one can identify the Majorana zero mode $\Psi$ with the parity of the bulk states, $\Psi|\Omega,\vec n\rangle=(-1)^{\sum_i n_i}|\Omega,\vec n\rangle$.
 Interactions can be viewed as inducing a hopping process from one Fock state to a neighboring one with a different energy and thus equilibration by interactions can be viewed as a delocalization process in Fock space as has been shown in the seminal work of Altshuler, Gefen, Kamenev, and Levitov \cite{Altshuler1997} on interacting chaotic quantum dots. 

 When wavefunctions are localized in Fock space, the Majorana zero mode will not decay and the plateau value of the $Y_r$ autocorrelation function, \eqref{Pdef},
is a number of order 1. In contrast, in the fully Fock space delocalized phase  a typical eigenstate will have the form $|\Psi_\text{typ}\rangle = \sum_{\vec n,\Omega} a_{\Omega\vec n}|\Omega,\vec n\rangle$ with $|a_{\Omega\vec n}|\sim 1/\sqrt{2^{L}}$, and the plateau value will be roughly
\begin{align}
    P_{\text{Fock-delocalized}}&\sim   |\langle \Psi_\text{typ}  | Y_r |\Psi_\text{typ}\rangle|^2  \nonumber \\
    &= \left|\sum_{\Omega,\vec{n},\vec{m}} a_{\Omega\vec n}^* a_{\Omega\vec m}    \langle \Omega,\vec n  | Y_r |\Omega,\vec m\rangle\right|^2 \nonumber \\
 &  \sim \left|\frac{{c}}{2^L}  \sum_{\vec{n}}  (\pm 1)  \right|^2 \sim \frac{1}{2^L},\label{eq:Pergodic} 
\end{align}
where we used that $Y_r$ and the Majorana mode $\Psi$ have a finite overlap, $Y_r \sim c \Psi$, that $\Psi$ has eigenvalues $\pm 1$, and that the sum of $N=2^L$ numbers of order $1$ with random signs is of order $\sqrt{N}$.
The value of the plateau which we observe numerically for small $J_x$ and $L$ is, however, much larger than that predicted by \eqref{eq:Pergodic}. A sizable plateau thus implies that the wavefunction remains at least partially localized in Fock space.  In this case the sum in \eqref{eq:Pergodic} runs only over a small number of terms, giving rise to large plateau values. The quantitative link of localization and size of the plateau is discussed numerically below.

To understand how $J_x$ induces a delocalization transition in a finite-size system, we need to consider three types of matrix elements of the interactions, $M_{\rm eM}$, $M_{\rm ee}$ and $M_{\rm ee}^b$. These describe the scattering of the boundary Majorana with 3 extended bulk states $M_{\rm eM}$, the momentum-conserving scattering of 4 extended bulk states $M_{\rm ee}$, and the scattering of 4 extended bulk states at the boundary without momentum conservation $M_{\rm ee}^b$.

A bulk state with momentum $k$ has a single-particle wavefunction of the form $\exp{(i k R_j)}/\sqrt{L}$ where $R_j$ denotes the particle position.
Thus a typical four-Majorana bulk-interaction matrix element in a clean system scales with $(J_x/\sqrt{L}^4)\sum_j \exp(i (k_1+k_2+k_3+k_4) R_j)=(J_x L/\sqrt{L}^4) \delta_{k_1+k_2+k_3+k_4,0}$. Each $k$ has $L$ possible values and there are therefore $L^3$ possible combinations of $k_i$ fulfilling momentum conservation. Thus 
bulk interactions connect a given Fock state to $Z_{\rm ee}=L^3$ other Fock states with a matrix element $ M_{\rm ee}$ of order $J_x L/\sqrt{L}^4$. 

We can repeat this argument for boundary scattering of extended bulk states. In this case the matrix element is suppressed by a factor $1/L$, $M^b_{\rm ee}\sim M_{\rm ee}/L$  but due to the absence of momentum conservation, the number of connecting neighbors in Fock space increases, $Z^b_{\rm ee}\sim Z_{\rm ee} L$.

Finally, the matrix element of a localized Majorana mode with wave-function $\phi_j$ interacting with three delocalized bulk modes scales with $(J_x/\sqrt{L}^3)\sum_j \phi_j \exp(i (k_1+k_2+k_3) R_j)\sim J_x/\sqrt{L}^3$, giving rise to a hopping to $ Z_{\rm eM} \sim L^3 $ neighbors in Fock space.

Thus, we have found the following scaling with system size $L$
\begin{align}
    M_{\rm eM} &\sim \frac{J_x}{(\sqrt{L})^3},\quad  & Z_{\rm eM} \sim L^3 \nonumber \\
    M_{\rm ee} &\sim \frac{J_x L}{(\sqrt{L})^4}=\frac{J_x}{L}, & Z_{\rm ee} \sim L^3 \nonumber \\
    M^b_{\rm ee} &\sim \frac{J_x}{(\sqrt{L})^4}, & Z^b_{\rm ee} \sim L^4 \label{eq:MZ}
\end{align}
Note that the $M_{\rm ee}$ and $M^b_{\rm ee}$ processes conserve the edge Majorana $\Psi$, or, equivalently the parity of bulk states $(-1)^{\sum_i n_i}$,  while the  $M_{\rm eM}$ processes flip $\Psi$. At least locally \cite{Altshuler1997}, the interaction induced scattering leads to a hopping problem on a Cayley tree \cite{Abou-Chacra_1973} with a large coordination number as sketched in Fig.~\ref{fig.flower}.

Importantly, the Fock states connected by the above described matrix elements typically have very different quasi-energies (defined modulo $2 \pi/T$). This difference of quasi-energies of states connected by $J_x$ depends on the chosen parameters.
For the following discussion, we focus on a regime where $g T/2$ and $J T/2$ are both numbers of order $1$ and where in the thermodynamic limit the Majorana zero mode is in the middle of a continuum spanned by the sums and differences of three bulk-state quasi-energies, $\epsilon_k$, i.e, one can find solutions for
$\epsilon_{k_1}\pm \epsilon_{k_2}\pm \epsilon_{k_3}=2 \pi n/T , n \in \mathbb Z$, in the thermodynamic limit.
A straightforward Fermi's Golden Rule argument \cite{HCY23} predicts that in this case the inverse lifetime of the Majorana scales with $J_x^2$, 
which is, e.g., fulfilled for the parameters chosen in Fig.~\ref{decay}.
Under these conditions, the typical differences of quasi-energies of Fock states connected by $J_x$ is of the order of $E_0 \sim J \sim g \sim 2 \pi/T$. 
If a given Fock state has $Z$ neighbors, the typical minimal energy difference is thus $E_0/Z$. Thus, one can obtain a first estimate of the delocalization transition by comparing $M_\alpha$ to $E_0/Z_\alpha$, where $\alpha$ labels the particular scattering process in \eqref{eq:MZ}. More precisely, it has been
shown by Abou-Chacra, Thouless, and Anderson \cite{Abou-Chacra_1973} that for large $Z$ and on a Cayley tree, the system starts to delocalize in Fock space for
\begin{align}
    M_\alpha \gtrsim \frac{E_0}{Z_\alpha \ln[Z_\alpha]}. \label{eq:transition}
\end{align}
Note that  the applicability of this formula for interaction-induced delocalization in Fock space has been controversial even for the most intensively studied problem of a chaotic interacting quantum dot, see Ref.~\cite{Gornyi2016} for a discussion of arguments and publications pointing to different dependencies on $Z_\alpha$, but large parts of the recent literature seem to agree with \eqref{eq:transition}, see e.g., \cite{Monteiro2021,Bulchandani22}. Our numerical results are at least consistent with \eqref{eq:transition} and we will use \eqref{eq:transition} only for qualitative arguments below.

Using \eqref{eq:transition} for the  processes of \eqref{eq:MZ} reveals that upon increasing $J_x$ starting from the non-interacting system, the delocalization of Fock states occurs first via the two bulk-mode scattering processes $M_{\rm ee}$ and $M_{\rm ee}^b$ at $J_x \sim  \DeltaE/L^2$ if we ignore logarithmic corrections. While $M_{\rm ee}^b \ll M_{\rm ee}$, it profits from a larger number of accessible states such that $M_{\rm ee} Z_{\rm ee}\sim M_{\rm ee}^b Z_{\rm ee}^b$. As nominally $\ln Z_{\rm ee}^b> \ln  Z_{\rm ee}$, \eqref{eq:transition} strictly speaking predicts that $M_{\rm ee}^b$ is more important than $M_{\rm ee}$ for very large systems. 

For our discussion, another effect is more important. Due to delocalization, one obtains an effective broadening of the quasi-energies of the Fock states  at $M_\alpha \sim  \DeltaE/Z_\alpha$ (ignoring again logarithms), of the order of $\DeltaE/Z_\alpha$ \cite{Monteiro2021} with $\DeltaE/Z_{\rm ee} \gg \DeltaE/Z_{\rm ee}^b$.
Thus, the broadening of Fock levels is dominated by the momentum-conserving bulk-scattering processes as we show numerically below.

This leads to the following physical picture upon increasing $J_x$ in a finite-size system: first, Fock space hopping induces a delocalization in Fock space at a small value of the  coupling, $J_x \sim \DeltaE/L^2$. This delocalization induces a broadening of the quasi-energies of Fock states, mainly driven by bulk-scattering processes shown as thick blue lines in Fig.~\ref{fig.flower}. This broadening effectively unblocks scattering processes 
involving the localized Majorana mode at the boundary for slightly larger $J_x$. Due to the `unblocking', the plateau in the $Y_r$ correlation function decays. We would like to stress that this analysis only applies to the case of a {\em single} zero-energy Majorana mode, which is the focus of this study. In the presence of several zero-energy modes there are extra degeneracies in the many-particle spectrum of the non-interacting system, which have to be taken into account separately.

To corroborate this picture numerically, we will show in the following
(i) that  the delocalization of bulk states in Fock space is governed by bulk scattering processes and (ii) 
that the localization/delocalization of Fock states governs the formation/decay of the plateau of the Majorana correlation function.

Denoting $|\beta_0\rangle$ as an eigenstate of the integrable system, or, equivalently a Fock state, our goal is to study the fate of the bulk conserved quantity  $|\beta_0\rangle\langle \beta_0|$ when integrability-breaking interactions are switched on. The infinite temperature autocorrelation function of this quantity is
$|\langle \beta_0| T_{\sigma, S/A}^n|\beta_0\rangle|^2$. We can also interpret this quantity as the return probability that the quantum system starting in the Fock state $|\beta_0\rangle$ is still in the same Fock state after $n$ time steps.
 Averaging the return probability over all conserved quantities gives
\begin{align}
 &   \text{Average Return Probability}(n) \nonumber\\
 &   = \frac{1}{2^L}\sum_{\beta_0}|\langle \beta_0| T_{\sigma, S/A}^n|\beta_0\rangle|^2\nonumber\\
 &= \frac{1}{2^L}\sum_{\beta_0, \beta,\beta'}
 |\langle \beta_0| \beta\rangle|^2
 |\langle\beta'|\beta_0\rangle|^2 e^{-i(\epsilon_{\beta}-\epsilon_{\beta'})T n}.\label{RPdef} 
\end{align}
Above, $\epsilon_{\beta}$ denote the quasi-energy associated with the eigenstate $|\beta\rangle$ of the non-integrable unitary $T_{\sigma, S/A}$. In \eqref{RPdef} we use the Fock states of the integrable model, $J_x=0$, thus neglecting the effect that Fock states will obtain Hartree-Fock style corrections from $J_x$ which may lead to a suppression of Fock state overlaps $\sim \exp[-{\rm constant} \, (J_x/\DeltaE)^2 L]$. Neglecting this is justified even for large $L$ as typical values of $J_x$ triggering Fock space delocalization are tiny, $J_x \sim \DeltaE/L^2$, according to \eqref{eq:transition}, see also Fig.~\ref{IPRfig} below.

The average return probability is plotted in Fig.~\ref{IPRTime} for the same parameters as Fig.~\ref{Jx0p1}. The grey lines correspond to taking the long time limit where one may project onto the diagonal ensemble $\beta=\beta'$. In this limit, the average return probability equals the 
inverse participation ratio (IPR) in Fock space
\begin{align}
&\text{Average Return Probability}(n\rightarrow \infty)\nonumber\\
&=\text{IPR}=\frac{1}{2^L} \sum_{\beta_0,\beta}  |\langle \beta_0 | \beta\rangle|^4\label{IRPdef}.
\end{align}
The IPR is a common measure of delocalization: a wave function delocalized over roughly $N$ states (here in Fock space) has  $|\langle \beta_0 | \beta\rangle|^2\sim 1/N$ for $N$ states $|\beta\rangle$, resulting in an IPR of order $N/N^2=1/N$ with $N\sim 2^L$ in the fully delocalized limit.

The Fermi's Golden Rule formula suggests that the inverse lifetime of a typical Fock state $|\beta\rangle$ in the delocalized phase (or for short times, see below) scales as $ 2 \pi \sum_{\beta'} |\langle \beta' | H_{\text{int}} |\beta \rangle|^2 \delta(\epsilon_\beta-\epsilon_\beta') \sim \sum_{\alpha} \Gamma_\alpha$ with $\Gamma_\alpha\sim|M_\alpha|^2 Z_\alpha/\DeltaE$, where we sum over all possible bulk, edge and Majorana scattering channels $\alpha$ and assume that the effective broadening of the $\delta$-function is larger than the effective level spacing of neighboring states in Fock space, $\DeltaE/Z_\alpha$. Using
\eqref{eq:MZ}, the shortest lifetime arises from bulk-scattering processes with $\Gamma_{\rm ee}\sim |M_{\rm ee}|^2 Z_{\rm ee}/E_0 \sim L J_x^2/E_0$. 
In contrast, the boundary scattering processes with and without the Majorana mode give $\Gamma_{\rm em}\sim \Gamma_{\rm ee}^b \sim J_x^2/E_0\ll \Gamma_{\rm ee}$ and can therefore be neglected.
Thus, we plot in Fig.~\ref{IPRTime} (lower panel) the average return probability as function of $J_x^2 n L $. For short times, we obtain an excellent collapse and the larger the system, the longer it follows the expected exponential decay on a time-scale of order $1/\Gamma_{\rm ee}$. This indicates that bulk scattering is -- as expected -- the dominant decay mechanism of Fock states.
The plateau obtained for larger times and small systems, which we identified with the IPR, implies that the wavefunctions remain partially localized in Fock space. Delocalization is expected to occur when the broadening of Fock levels, $\Gamma_{\rm ee}$, becomes larger than the relevant level spacing, $\Gamma_{\rm ee} \gtrsim \DeltaE/Z_{\rm ee}$, or, equivalently, $M_{\rm ee}\gtrsim \DeltaE/Z_{\rm ee}$.  This condition is (up to the logarithm not covered by this simple argument) consistent with \eqref{eq:transition}.

\begin{figure}
\includegraphics[width = 0.45\textwidth]{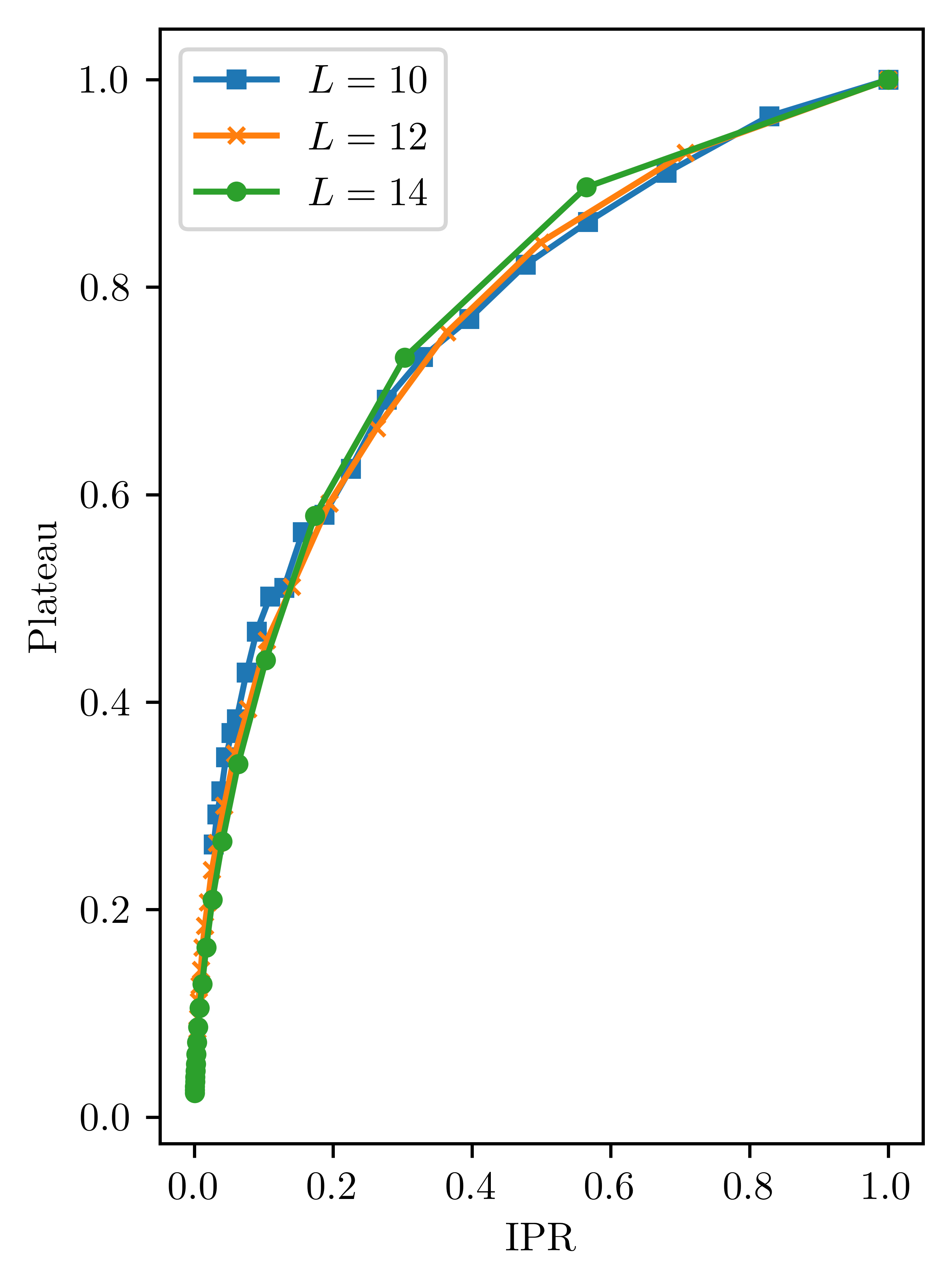}
\caption{ Normalized Plateau $P(J_x)/P(J_x=0)$ vs IPR for the case where there is no $Z_2$ symmetry and $JT/2=1.0, T=2.0,g=0.3$.}\label{IPR-P}
\end{figure}

After we have established that the delocalization in Fock space is governed by bulk scattering processes,
we will now show that this delocalization governs also the destruction of the plateau of the Majorana correlation function. Fig.~\ref{IPRfig} plots for three different system sizes both the plateau heights  and the average Fock space IPR, which is used to keep track of delocalization in Fock space (see discussion above), as a function of the interaction strength $J_x$. As expected, both quantities vanish for large $J_x$ when the system becomes ergodic. For larger systems, smaller values of $J_x$ are sufficient to induce both delocalization and a suppression of the plateau. To show that these two processes, delocalization by bulk scattering and decay of the Majorana mode, are directly linked with each other, we plot in Fig.~\ref{IPR-P} the (normalized) plateau value as a function of the IPR, using the data of Fig.~\ref{IPRfig}. 
Remarkably, different system sizes now approximately collapse on a single curve. It is a surprising numerical observation that the plateau height as a function of the IPR appears to depend only weakly on system size. The figure is fully consistent with our theoretical scenario that the delocalization in Fock space, measured by the reduction of the IPR, activates edge-Majorana scattering and
suppresses therefore the plateau in the edge-correlation function.  In our example, the plateau is reduced to half of its size roughly when the IPR is $0.1$ implying a delocalization in Fock space covering roughly $10$ Fock states. 

\begin{figure}
\includegraphics[width = 0.45\textwidth]{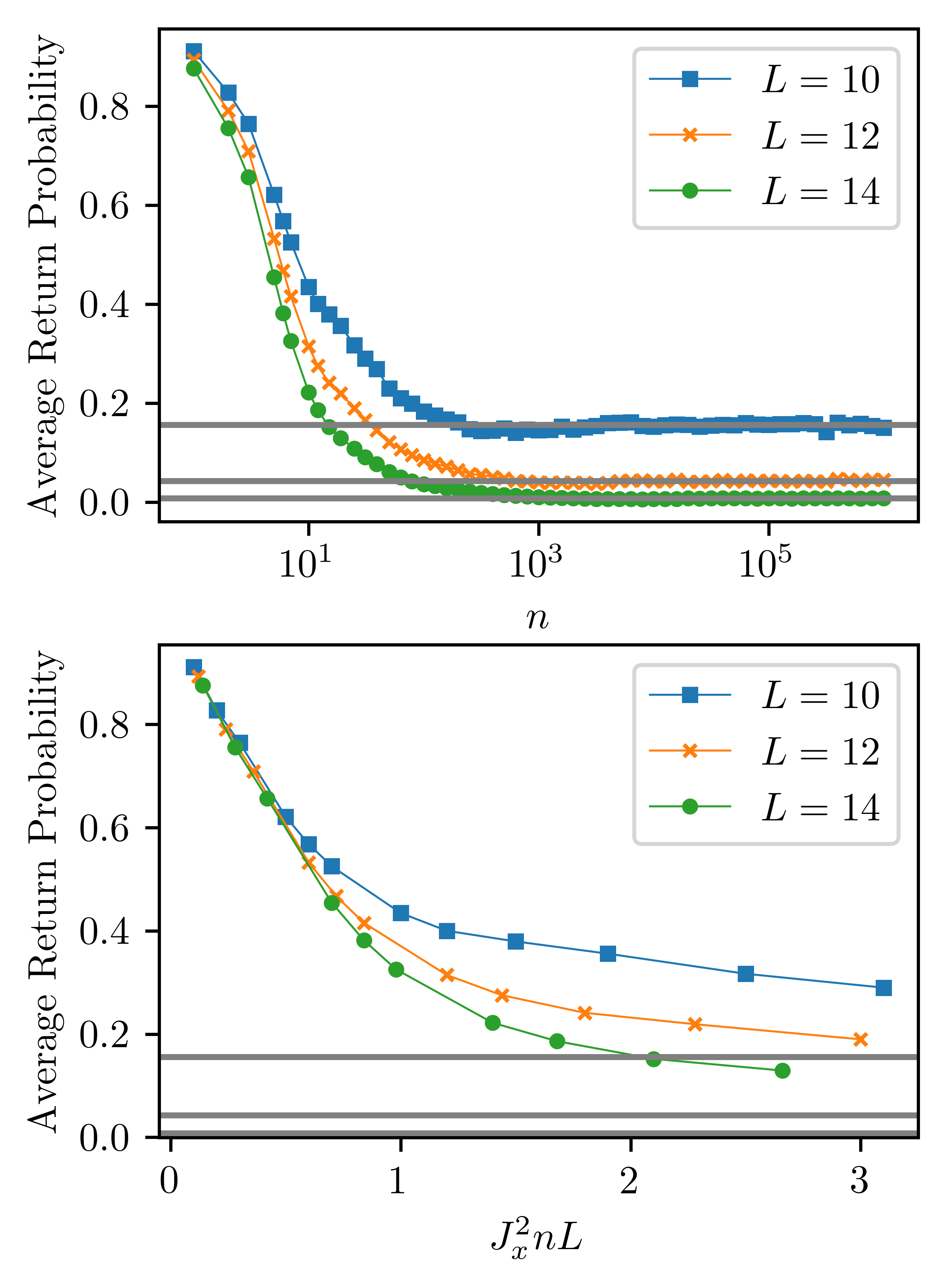}
\caption{\label{IPRTime} Top panel: Average Return Probability vs stroboscopic time $n$. After an initial transient, this quantity approaches a plateau whose magnitude agrees well with the IPR, the latter indicated  by the grey lines. 
Bottom panel: Same as top panel but now plotted vs $J_x^2 n L$.
The parameters for both panels are $JT/2=1.0, T=2.0, gT/2=0.3$ and $J_x =0.1$, with no $Z_2$ symmetry.}
\end{figure}

\begin{figure}
\includegraphics[width = 0.45\textwidth]{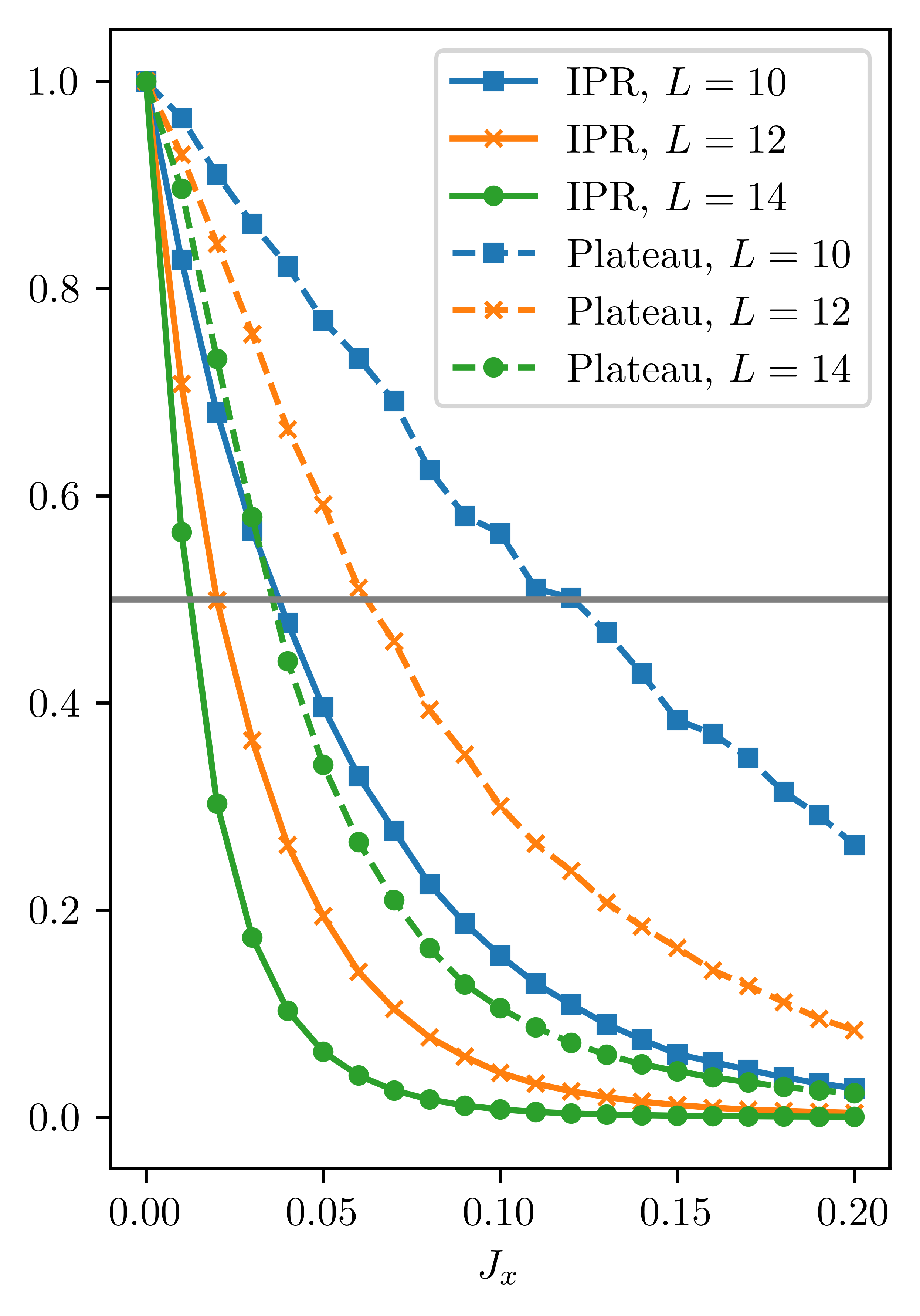}
\caption{IPR and normalized plateau heights $P(J_x)/P(J_x=0)$ vs $J_x$ for the case where there is no $Z_2$ symmetry and $JT/2=1.0, T=2.0,g=0.3$.}\label{IPRfig}
\end{figure}

\section{Conclusions} \label{sec:Conc}
Interacting Floquet systems are a bedrock for studying nonequilibrium and topological phenomena, where the latter are sensitive to boundary conditions. 
In this paper, we have explored the role of a special boundary condition,
the duality twisted boundary condition on these systems. The role of the duality twist was to introduce a single Majorana zero mode in an integrable Floquet unitary, for all coupling parameters of the unitary. 
In addition, this isolated Majorana zero mode did not lead to any degeneracies of the Floquet eigenspectra. This is quite unlike
open boundary conditions where the Majorana zero modes appear in pairs, arise for special couplings (transverse fields smaller than
Ising interactions), and cause a degeneracy of the Floquet spectra. We showed that, even though for spatially uniform transverse fields and Ising couplings,
the Majorana zero mode is localized at the duality twist, this is actually a special limit where a domain wall is absent. 
In particular, a unitary transformation can be performed that moves the
duality twist relative to the Majorana zero mode, with the Majorana zero mode now residing at a domain wall separating 
two regions that are related by the interchange of the Ising couplings and the transverse field couplings.   

We studied the effect of interactions for the case where the Majorana zero mode resides near the duality twist, i.e., in the absence of a domain wall. 
While the Majorana zero mode does not survive in the thermodynamic limit when interactions are non-zero, nevertheless, we found that
for finite system sizes, the Majorana zero mode is related to an emergent symmetry. The signature of this was an autocorrelation function which after an initial transient, approached a plateau. 
The height of the plateau was found to agree very well with a numerically constructed conserved quantity which overlaps with the integrable Majorana zero mode. 
These results were quite insensitive to whether the interactions preserved the $Z_2$ symmetry or not, with differences appearing only in details such as, the
precise values of the plateau heights. 

A theory was presented for the eventual decay of the plateau in the thermodynamic limit in the language of Fock space delocalization \cite{Bulchandani22}. In the absence of interactions,
the conserved quantities, both boundary and bulk, are localized in Fock space. Once interactions are switched on, hopping in Fock space sets in,  but for very small interactions the many-particle wavefunctions remain localized in Fock space. We showed that the matrix elements for decay of the bulk conserved quantities are larger than the boundary conserved quantity. Thus as the interactions are increased, the wavefunction starts to delocalize in Fock space and bulk
conserved quantities decay first, followed by the boundary conserved quantity. This phenomena was supported by ED which showed that the average autocorrelation
function of all bulk conserved quantities reached a steady-state at earlier stroboscopic times than the boundary autocorrelation function, where the
steady-state value of the former is the IPR. In addition, we showed that the IPR decay directly controls the decay of the  plateau height associated with the boundary Majorana mode.  We  expect that the physics of a multi-step localization/delocalization transition will also have applications for other integrable systems subject to weak integrability-breaking perturbations.

There are many open questions. Generalizing the study of the physics of duality twisted boundary conditions to other systems, such as $Z_N$ spin chains is an obvious direction of study. Further, the role of more than one topological defect on the dynamics of the system, is an interesting direction of research.

{\sl Acknowledgements:} The authors are deeply indebted to Alex Altland and Dries Sels for many helpful discussions. 
This work was supported by the US National Science Foundation under Grant NSF-DMR 2018358 (AM), in part by the US National Science Foundation under Grant No. NSF PHY-1748958 (AM) and by the German Research Foundation within
CRC183 (project number 277101999, subproject A01 (AR) and a Mercator fellowship (AM)).

\newpage

\begin{thebibliography}{36}%
\makeatletter
\providecommand \@ifxundefined [1]{%
 \@ifx{#1\undefined}
}%
\providecommand \@ifnum [1]{%
 \ifnum #1\expandafter \@firstoftwo
 \else \expandafter \@secondoftwo
 \fi
}%
\providecommand \@ifx [1]{%
 \ifx #1\expandafter \@firstoftwo
 \else \expandafter \@secondoftwo
 \fi
}%
\providecommand \natexlab [1]{#1}%
\providecommand \enquote  [1]{``#1''}%
\providecommand \bibnamefont  [1]{#1}%
\providecommand \bibfnamefont [1]{#1}%
\providecommand \citenamefont [1]{#1}%
\providecommand \href@noop [0]{\@secondoftwo}%
\providecommand \href [0]{\begingroup \@sanitize@url \@href}%
\providecommand \@href[1]{\@@startlink{#1}\@@href}%
\providecommand \@@href[1]{\endgroup#1\@@endlink}%
\providecommand \@sanitize@url [0]{\catcode `\\12\catcode `\$12\catcode
  `\&12\catcode `\#12\catcode `\^12\catcode `\_12\catcode `\%12\relax}%
\providecommand \@@startlink[1]{}%
\providecommand \@@endlink[0]{}%
\providecommand \url  [0]{\begingroup\@sanitize@url \@url }%
\providecommand \@url [1]{\endgroup\@href {#1}{\urlprefix }}%
\providecommand \urlprefix  [0]{URL }%
\providecommand \Eprint [0]{\href }%
\providecommand \doibase [0]{https://doi.org/}%
\providecommand \selectlanguage [0]{\@gobble}%
\providecommand \bibinfo  [0]{\@secondoftwo}%
\providecommand \bibfield  [0]{\@secondoftwo}%
\providecommand \translation [1]{[#1]}%
\providecommand \BibitemOpen [0]{}%
\providecommand \bibitemStop [0]{}%
\providecommand \bibitemNoStop [0]{.\EOS\space}%
\providecommand \EOS [0]{\spacefactor3000\relax}%
\providecommand \BibitemShut  [1]{\csname bibitem#1\endcsname}%
\let\auto@bib@innerbib\@empty
\bibitem [{\citenamefont {Kramers}\ and\ \citenamefont
  {Wannier}(1941)}]{Kramers1941}%
  \BibitemOpen
  \bibfield  {author} {\bibinfo {author} {\bibfnamefont {H.~A.}\ \bibnamefont
  {Kramers}}\ and\ \bibinfo {author} {\bibfnamefont {G.~H.}\ \bibnamefont
  {Wannier}},\ }\bibfield  {title} {\bibinfo {title} {{Statistics of the
  two-dimensional ferromagnet. Part 1.}},\ }\href
  {https://doi.org/10.1103/PhysRev.60.252} {\bibfield  {journal} {\bibinfo
  {journal} {Phys. Rev.}\ }\textbf {\bibinfo {volume} {60}},\ \bibinfo {pages}
  {252} (\bibinfo {year} {1941})}\BibitemShut {NoStop}%
\bibitem [{\citenamefont {Aasen}\ \emph {et~al.}(2016)\citenamefont {Aasen},
  \citenamefont {Mong},\ and\ \citenamefont {Fendley}}]{Fendley16}%
  \BibitemOpen
  \bibfield  {author} {\bibinfo {author} {\bibfnamefont {D.}~\bibnamefont
  {Aasen}}, \bibinfo {author} {\bibfnamefont {R.~S.~K.}\ \bibnamefont {Mong}},\
  and\ \bibinfo {author} {\bibfnamefont {P.}~\bibnamefont {Fendley}},\
  }\bibfield  {title} {\bibinfo {title} {Topological defects on the lattice: I.
  the ising model},\ }\href {https://doi.org/10.1088/1751-8113/49/35/354001}
  {\bibfield  {journal} {\bibinfo  {journal} {Journal of Physics A:
  Mathematical and Theoretical}\ }\textbf {\bibinfo {volume} {49}},\ \bibinfo
  {pages} {354001} (\bibinfo {year} {2016})}\BibitemShut {NoStop}%
\bibitem [{\citenamefont {Aasen}\ \emph {et~al.}(2020)\citenamefont {Aasen},
  \citenamefont {Fendley},\ and\ \citenamefont {Mong}}]{Fendley20}%
  \BibitemOpen
  \bibfield  {author} {\bibinfo {author} {\bibfnamefont {D.}~\bibnamefont
  {Aasen}}, \bibinfo {author} {\bibfnamefont {P.}~\bibnamefont {Fendley}},\
  and\ \bibinfo {author} {\bibfnamefont {R.~S.}\ \bibnamefont {Mong}},\
  }\bibfield  {title} {\bibinfo {title} {Topological defects on the lattice:
  dualities and degeneracies},\ }\href@noop {} {\bibfield  {journal} {\bibinfo
  {journal} {arXiv preprint arXiv:2008.08598}\ } (\bibinfo {year}
  {2020})}\BibitemShut {NoStop}%
\bibitem [{\citenamefont {Lootens}\ \emph {et~al.}(9091)\citenamefont
  {Lootens}, \citenamefont {Delcamp}, \citenamefont {Ortiz},\ and\
  \citenamefont {Verstraete}}]{Verstraete21}%
  \BibitemOpen
  \bibfield  {author} {\bibinfo {author} {\bibfnamefont {L.}~\bibnamefont
  {Lootens}}, \bibinfo {author} {\bibfnamefont {C.}~\bibnamefont {Delcamp}},
  \bibinfo {author} {\bibfnamefont {G.}~\bibnamefont {Ortiz}},\ and\ \bibinfo
  {author} {\bibfnamefont {F.}~\bibnamefont {Verstraete}},\ }\href@noop {}
  {\bibinfo {title} {Dualities in one-dimensional quantum lattice models:
  symmetric hamiltonians and matrix product operator intertwiners}} (\bibinfo
  {year} {arXiv::2112.09091})\BibitemShut {NoStop}%
\bibitem [{\citenamefont {Verlinde}(1988)}]{Verlinde:1988sn}%
  \BibitemOpen
  \bibfield  {author} {\bibinfo {author} {\bibfnamefont {E.~P.}\ \bibnamefont
  {Verlinde}},\ }\bibfield  {title} {\bibinfo {title} {{Fusion Rules and
  Modular Transformations in 2D Conformal Field Theory}},\ }\href
  {https://doi.org/10.1016/0550-3213(88)90603-7} {\bibfield  {journal}
  {\bibinfo  {journal} {Nucl. Phys. B}\ }\textbf {\bibinfo {volume} {300}},\
  \bibinfo {pages} {360} (\bibinfo {year} {1988})}\BibitemShut {NoStop}%
\bibitem [{\citenamefont {Cardy}(1989)}]{CARDY1989581}%
  \BibitemOpen
  \bibfield  {author} {\bibinfo {author} {\bibfnamefont {J.~L.}\ \bibnamefont
  {Cardy}},\ }\bibfield  {title} {\bibinfo {title} {Boundary conditions, fusion
  rules and the verlinde formula},\ }\href
  {https://doi.org/https://doi.org/10.1016/0550-3213(89)90521-X} {\bibfield
  {journal} {\bibinfo  {journal} {Nuclear Physics B}\ }\textbf {\bibinfo
  {volume} {324}},\ \bibinfo {pages} {581} (\bibinfo {year}
  {1989})}\BibitemShut {NoStop}%
\bibitem [{\citenamefont {Petkova}\ and\ \citenamefont
  {Zuber}(2001)}]{Petkova00}%
  \BibitemOpen
  \bibfield  {author} {\bibinfo {author} {\bibfnamefont {V.~B.}\ \bibnamefont
  {Petkova}}\ and\ \bibinfo {author} {\bibfnamefont {J.~B.}\ \bibnamefont
  {Zuber}},\ }\bibfield  {title} {\bibinfo {title} {{Generalized twisted
  partition functions}},\ }\href
  {https://doi.org/10.1016/S0370-2693(01)00276-3} {\bibfield  {journal}
  {\bibinfo  {journal} {Phys. Lett. B}\ }\textbf {\bibinfo {volume} {504}},\
  \bibinfo {pages} {157} (\bibinfo {year} {2001})},\ \Eprint
  {https://arxiv.org/abs/hep-th/0011021} {arXiv:hep-th/0011021} \BibitemShut
  {NoStop}%
\bibitem [{\citenamefont {Fr\"ohlich}\ \emph {et~al.}(2004)\citenamefont
  {Fr\"ohlich}, \citenamefont {Fuchs}, \citenamefont {Runkel},\ and\
  \citenamefont {Schweigert}}]{FrohlichFucksRunkelSchweigert}%
  \BibitemOpen
  \bibfield  {author} {\bibinfo {author} {\bibfnamefont {J.}~\bibnamefont
  {Fr\"ohlich}}, \bibinfo {author} {\bibfnamefont {J.}~\bibnamefont {Fuchs}},
  \bibinfo {author} {\bibfnamefont {I.}~\bibnamefont {Runkel}},\ and\ \bibinfo
  {author} {\bibfnamefont {C.}~\bibnamefont {Schweigert}},\ }\bibfield  {title}
  {\bibinfo {title} {Kramers-wannier duality from conformal defects},\ }\href
  {https://doi.org/10.1103/PhysRevLett.93.070601} {\bibfield  {journal}
  {\bibinfo  {journal} {Phys. Rev. Lett.}\ }\textbf {\bibinfo {volume} {93}},\
  \bibinfo {pages} {070601} (\bibinfo {year} {2004})}\BibitemShut {NoStop}%
\bibitem [{\citenamefont {Chang}\ \emph {et~al.}(2019)\citenamefont {Chang},
  \citenamefont {Lin}, \citenamefont {Shao}, \citenamefont {Wang},\ and\
  \citenamefont {Yin}}]{Chang:2018iay}%
  \BibitemOpen
  \bibfield  {author} {\bibinfo {author} {\bibfnamefont {C.-M.}\ \bibnamefont
  {Chang}}, \bibinfo {author} {\bibfnamefont {Y.-H.}\ \bibnamefont {Lin}},
  \bibinfo {author} {\bibfnamefont {S.-H.}\ \bibnamefont {Shao}}, \bibinfo
  {author} {\bibfnamefont {Y.}~\bibnamefont {Wang}},\ and\ \bibinfo {author}
  {\bibfnamefont {X.}~\bibnamefont {Yin}},\ }\bibfield  {title} {\bibinfo
  {title} {{Topological Defect Lines and Renormalization Group Flows in Two
  Dimensions}},\ }\href {https://doi.org/10.1007/JHEP01(2019)026} {\bibfield
  {journal} {\bibinfo  {journal} {JHEP}\ }\textbf {\bibinfo {volume} {01}},\
  \bibinfo {pages} {026}},\ \Eprint {https://arxiv.org/abs/1802.04445}
  {arXiv:1802.04445 [hep-th]} \BibitemShut {NoStop}%
\bibitem [{\citenamefont {Tan}\ \emph {et~al.}(2022)\citenamefont {Tan},
  \citenamefont {Wang},\ and\ \citenamefont {Mitra}}]{Tan22}%
  \BibitemOpen
  \bibfield  {author} {\bibinfo {author} {\bibfnamefont {M.~T.}\ \bibnamefont
  {Tan}}, \bibinfo {author} {\bibfnamefont {Y.}~\bibnamefont {Wang}},\ and\
  \bibinfo {author} {\bibfnamefont {A.}~\bibnamefont {Mitra}},\ }\bibfield
  {title} {\bibinfo {title} {Topological defects in floquet circuits},\
  }\href@noop {} {\bibfield  {journal} {\bibinfo  {journal} {arXiv:2206.06272}\
  } (\bibinfo {year} {2022})}\BibitemShut {NoStop}%
\bibitem [{\citenamefont {Tantivasadakarn}\ \emph {et~al.}(2021)\citenamefont
  {Tantivasadakarn}, \citenamefont {Thorngren}, \citenamefont {Vishwanath},\
  and\ \citenamefont {Verresen}}]{Verresen21}%
  \BibitemOpen
  \bibfield  {author} {\bibinfo {author} {\bibfnamefont {N.}~\bibnamefont
  {Tantivasadakarn}}, \bibinfo {author} {\bibfnamefont {R.}~\bibnamefont
  {Thorngren}}, \bibinfo {author} {\bibfnamefont {A.}~\bibnamefont
  {Vishwanath}},\ and\ \bibinfo {author} {\bibfnamefont {R.}~\bibnamefont
  {Verresen}},\ }\bibfield  {title} {\bibinfo {title} {Long-range entanglement
  from measuring symmetry-protected topological phases},\ }\href@noop {}
  {\bibfield  {journal} {\bibinfo  {journal} {arXiv:2112.01519}\ } (\bibinfo
  {year} {2021})}\BibitemShut {NoStop}%
\bibitem [{\citenamefont {Aasen}\ \emph {et~al.}(2022)\citenamefont {Aasen},
  \citenamefont {Wang},\ and\ \citenamefont {Hastings}}]{Aasen_2022}%
  \BibitemOpen
  \bibfield  {author} {\bibinfo {author} {\bibfnamefont {D.}~\bibnamefont
  {Aasen}}, \bibinfo {author} {\bibfnamefont {Z.}~\bibnamefont {Wang}},\ and\
  \bibinfo {author} {\bibfnamefont {M.~B.}\ \bibnamefont {Hastings}},\
  }\bibfield  {title} {\bibinfo {title} {Adiabatic paths of hamiltonians,
  symmetries of topological order, and automorphism codes},\ }\bibfield
  {journal} {\bibinfo  {journal} {Physical Review B}\ }\textbf {\bibinfo
  {volume} {106}},\ \href {https://doi.org/10.1103/physrevb.106.085122}
  {10.1103/physrevb.106.085122} (\bibinfo {year} {2022})\BibitemShut {NoStop}%
\bibitem [{\citenamefont {Levy}(1991)}]{Levy91}%
  \BibitemOpen
  \bibfield  {author} {\bibinfo {author} {\bibfnamefont {D.}~\bibnamefont
  {Levy}},\ }\bibfield  {title} {\bibinfo {title} {Algebraic structure of
  translation-invariant spin-1/2 xxz and q-potts quantum chains},\ }\href@noop
  {} {\bibfield  {journal} {\bibinfo  {journal} {Phys. Rev. Lett.}\ }\textbf
  {\bibinfo {volume} {67}},\ \bibinfo {pages} {1971} (\bibinfo {year}
  {1991})}\BibitemShut {NoStop}%
\bibitem [{\citenamefont {Grimm}\ and\ \citenamefont
  {Schutz}(1993)}]{Schutz93}%
  \BibitemOpen
  \bibfield  {author} {\bibinfo {author} {\bibfnamefont {U.}~\bibnamefont
  {Grimm}}\ and\ \bibinfo {author} {\bibfnamefont {G.}~\bibnamefont {Schutz}},\
  }\bibfield  {title} {\bibinfo {title} {The spin-1/2xxz heisenberg chain, the
  quantum algebra $u_{q}\left[sl(2)\right]$, and duality transformations for
  minimal models},\ }\href@noop {} {\bibfield  {journal} {\bibinfo  {journal}
  {J. Stat. Phys.}\ }\textbf {\bibinfo {volume} {71}},\ \bibinfo {pages} {923}
  (\bibinfo {year} {1993})}\BibitemShut {NoStop}%
\bibitem [{\citenamefont {Oshikawa}\ and\ \citenamefont
  {Affleck}(1997)}]{Oshikawa97}%
  \BibitemOpen
  \bibfield  {author} {\bibinfo {author} {\bibfnamefont {M.}~\bibnamefont
  {Oshikawa}}\ and\ \bibinfo {author} {\bibfnamefont {I.}~\bibnamefont
  {Affleck}},\ }\bibfield  {title} {\bibinfo {title} {Boundary conformal field
  theory approach to the critical two-dimensional ising model with a defect
  line},\ }\href
  {https://doi.org/https://doi.org/10.1016/S0550-3213(97)00219-8} {\bibfield
  {journal} {\bibinfo  {journal} {Nuclear Physics B}\ }\textbf {\bibinfo
  {volume} {495}},\ \bibinfo {pages} {533} (\bibinfo {year}
  {1997})}\BibitemShut {NoStop}%
\bibitem [{\citenamefont {Chui}\ \emph {et~al.}(2001)\citenamefont {Chui},
  \citenamefont {Mercat}, \citenamefont {Orrick},\ and\ \citenamefont
  {Pearce}}]{Pearce01}%
  \BibitemOpen
  \bibfield  {author} {\bibinfo {author} {\bibfnamefont {C.}~\bibnamefont
  {Chui}}, \bibinfo {author} {\bibfnamefont {C.}~\bibnamefont {Mercat}},
  \bibinfo {author} {\bibfnamefont {W.~P.}\ \bibnamefont {Orrick}},\ and\
  \bibinfo {author} {\bibfnamefont {P.~A.}\ \bibnamefont {Pearce}},\ }\bibfield
   {title} {\bibinfo {title} {Integrable lattice realizations of conformal
  twisted boundary conditions},\ }\href@noop {} {\bibfield  {journal} {\bibinfo
   {journal} {Physics Letters B}\ }\textbf {\bibinfo {volume} {517}},\ \bibinfo
  {pages} {429} (\bibinfo {year} {2001})}\BibitemShut {NoStop}%
\bibitem [{\citenamefont {Grimm}(2002)}]{Grimm:2001dr}%
  \BibitemOpen
  \bibfield  {author} {\bibinfo {author} {\bibfnamefont {U.}~\bibnamefont
  {Grimm}},\ }\bibfield  {title} {\bibinfo {title} {{Spectrum of a duality
  twisted Ising quantum chain}},\ }\href
  {https://doi.org/10.1088/0305-4470/35/3/101} {\bibfield  {journal} {\bibinfo
  {journal} {J. Phys. A}\ }\textbf {\bibinfo {volume} {35}},\ \bibinfo {pages}
  {L25} (\bibinfo {year} {2002})},\ \Eprint
  {https://arxiv.org/abs/hep-th/0111157} {arXiv:hep-th/0111157} \BibitemShut
  {NoStop}%
\bibitem [{\citenamefont {Roy}\ and\ \citenamefont {Saleur}(7927)}]{Saleur21}%
  \BibitemOpen
  \bibfield  {author} {\bibinfo {author} {\bibfnamefont {A.}~\bibnamefont
  {Roy}}\ and\ \bibinfo {author} {\bibfnamefont {H.}~\bibnamefont {Saleur}},\
  }\href {https://doi.org/10.48550/ARXIV.2111.07927} {\bibinfo {title}
  {Entanglement entropy in critical quantum spin chains with boundaries and
  defects}} (\bibinfo {year} {arXiv:2111.07927})\BibitemShut {NoStop}%
\bibitem [{\citenamefont {Fendley}(2016)}]{FendleyXYZ}%
  \BibitemOpen
  \bibfield  {author} {\bibinfo {author} {\bibfnamefont {P.}~\bibnamefont
  {Fendley}},\ }\bibfield  {title} {\bibinfo {title} {Strong zero modes and
  eigenstate phase transitions in the xyz/interacting majorana chain},\ }\href
  {https://doi.org/10.1088/1751-8113/49/30/30LT01} {\bibfield  {journal}
  {\bibinfo  {journal} {Journal of Physics A: Mathematical and Theoretical}\
  }\textbf {\bibinfo {volume} {49}},\ \bibinfo {pages} {30LT01} (\bibinfo
  {year} {2016})}\BibitemShut {NoStop}%
\bibitem [{\citenamefont {Thakurathi}\ \emph {et~al.}(2013)\citenamefont
  {Thakurathi}, \citenamefont {Patel}, \citenamefont {Sen},\ and\ \citenamefont
  {Dutta}}]{Sen13}%
  \BibitemOpen
  \bibfield  {author} {\bibinfo {author} {\bibfnamefont {M.}~\bibnamefont
  {Thakurathi}}, \bibinfo {author} {\bibfnamefont {A.~A.}\ \bibnamefont
  {Patel}}, \bibinfo {author} {\bibfnamefont {D.}~\bibnamefont {Sen}},\ and\
  \bibinfo {author} {\bibfnamefont {A.}~\bibnamefont {Dutta}},\ }\bibfield
  {title} {\bibinfo {title} {Floquet generation of majorana end modes and
  topological invariants},\ }\href {https://doi.org/10.1103/PhysRevB.88.155133}
  {\bibfield  {journal} {\bibinfo  {journal} {Phys. Rev. B}\ }\textbf {\bibinfo
  {volume} {88}},\ \bibinfo {pages} {155133} (\bibinfo {year}
  {2013})}\BibitemShut {NoStop}%
\bibitem [{\citenamefont {Yates}\ \emph {et~al.}(2019)\citenamefont {Yates},
  \citenamefont {Essler},\ and\ \citenamefont {Mitra}}]{Yates19}%
  \BibitemOpen
  \bibfield  {author} {\bibinfo {author} {\bibfnamefont {D.~J.}\ \bibnamefont
  {Yates}}, \bibinfo {author} {\bibfnamefont {F.~H.~L.}\ \bibnamefont
  {Essler}},\ and\ \bibinfo {author} {\bibfnamefont {A.}~\bibnamefont
  {Mitra}},\ }\bibfield  {title} {\bibinfo {title} {Almost strong
  ($0,\ensuremath{\pi}$) edge modes in clean interacting one-dimensional
  floquet systems},\ }\href {https://doi.org/10.1103/PhysRevB.99.205419}
  {\bibfield  {journal} {\bibinfo  {journal} {Phys. Rev. B}\ }\textbf {\bibinfo
  {volume} {99}},\ \bibinfo {pages} {205419} (\bibinfo {year}
  {2019})}\BibitemShut {NoStop}%
\bibitem [{\citenamefont {Hauru}\ \emph {et~al.}(2016)\citenamefont {Hauru},
  \citenamefont {Evenbly}, \citenamefont {Ho}, \citenamefont {Gaiotto},\ and\
  \citenamefont {Vidal}}]{Vidal16}%
  \BibitemOpen
  \bibfield  {author} {\bibinfo {author} {\bibfnamefont {M.}~\bibnamefont
  {Hauru}}, \bibinfo {author} {\bibfnamefont {G.}~\bibnamefont {Evenbly}},
  \bibinfo {author} {\bibfnamefont {W.~W.}\ \bibnamefont {Ho}}, \bibinfo
  {author} {\bibfnamefont {D.}~\bibnamefont {Gaiotto}},\ and\ \bibinfo {author}
  {\bibfnamefont {G.}~\bibnamefont {Vidal}},\ }\bibfield  {title} {\bibinfo
  {title} {Topological conformal defects with tensor networks},\ }\href
  {https://doi.org/10.1103/PhysRevB.94.115125} {\bibfield  {journal} {\bibinfo
  {journal} {Phys. Rev. B}\ }\textbf {\bibinfo {volume} {94}},\ \bibinfo
  {pages} {115125} (\bibinfo {year} {2016})}\BibitemShut {NoStop}%
\bibitem [{\citenamefont {Yeh}\ \emph {et~al.}(2023)\citenamefont {Yeh},
  \citenamefont {Rosch},\ and\ \citenamefont {Mitra}}]{HCY23}%
  \BibitemOpen
  \bibfield  {author} {\bibinfo {author} {\bibfnamefont {H.-C.}\ \bibnamefont
  {Yeh}}, \bibinfo {author} {\bibfnamefont {A.}~\bibnamefont {Rosch}},\ and\
  \bibinfo {author} {\bibfnamefont {A.}~\bibnamefont {Mitra}},\ }\bibfield
  {title} {\bibinfo {title} {Decay rates of almost strong modes in floquet spin
  chains beyond fermi's golden rule},\ }\href@noop {} {\bibfield  {journal}
  {\bibinfo  {journal} {(in preparation)}\ } (\bibinfo {year}
  {2023})}\BibitemShut {NoStop}%
\bibitem [{\citenamefont {Rabson}\ \emph {et~al.}(2004)\citenamefont {Rabson},
  \citenamefont {Narozhny},\ and\ \citenamefont {Millis}}]{Rabson04}%
  \BibitemOpen
  \bibfield  {author} {\bibinfo {author} {\bibfnamefont {D.~A.}\ \bibnamefont
  {Rabson}}, \bibinfo {author} {\bibfnamefont {B.~N.}\ \bibnamefont
  {Narozhny}},\ and\ \bibinfo {author} {\bibfnamefont {A.~J.}\ \bibnamefont
  {Millis}},\ }\bibfield  {title} {\bibinfo {title} {Crossover from poisson to
  wigner-dyson level statistics in spin chains with integrability breaking},\
  }\href {https://doi.org/10.1103/PhysRevB.69.054403} {\bibfield  {journal}
  {\bibinfo  {journal} {Phys. Rev. B}\ }\textbf {\bibinfo {volume} {69}},\
  \bibinfo {pages} {054403} (\bibinfo {year} {2004})}\BibitemShut {NoStop}%
\bibitem [{\citenamefont {Jung}\ \emph {et~al.}(2006)\citenamefont {Jung},
  \citenamefont {Helmes},\ and\ \citenamefont {Rosch}}]{Jung2006}%
  \BibitemOpen
  \bibfield  {author} {\bibinfo {author} {\bibfnamefont {P.}~\bibnamefont
  {Jung}}, \bibinfo {author} {\bibfnamefont {R.~W.}\ \bibnamefont {Helmes}},\
  and\ \bibinfo {author} {\bibfnamefont {A.}~\bibnamefont {Rosch}},\ }\bibfield
   {title} {\bibinfo {title} {Transport in almost integrable models: Perturbed
  heisenberg chains},\ }\href {https://doi.org/10.1103/PhysRevLett.96.067202}
  {\bibfield  {journal} {\bibinfo  {journal} {Phys. Rev. Lett.}\ }\textbf
  {\bibinfo {volume} {96}},\ \bibinfo {pages} {067202} (\bibinfo {year}
  {2006})}\BibitemShut {NoStop}%
\bibitem [{\citenamefont {Jung}\ and\ \citenamefont {Rosch}(2007)}]{Jung2007}%
  \BibitemOpen
  \bibfield  {author} {\bibinfo {author} {\bibfnamefont {P.}~\bibnamefont
  {Jung}}\ and\ \bibinfo {author} {\bibfnamefont {A.}~\bibnamefont {Rosch}},\
  }\bibfield  {title} {\bibinfo {title} {Spin conductivity in almost integrable
  spin chains},\ }\href {https://doi.org/10.1103/PhysRevB.76.245108} {\bibfield
   {journal} {\bibinfo  {journal} {Phys. Rev. B}\ }\textbf {\bibinfo {volume}
  {76}},\ \bibinfo {pages} {245108} (\bibinfo {year} {2007})}\BibitemShut
  {NoStop}%
\bibitem [{\citenamefont {Rigol}\ \emph {et~al.}(2008)\citenamefont {Rigol},
  \citenamefont {Dunjko},\ and\ \citenamefont {Olshanii}}]{Rigol2008}%
  \BibitemOpen
  \bibfield  {author} {\bibinfo {author} {\bibfnamefont {M.}~\bibnamefont
  {Rigol}}, \bibinfo {author} {\bibfnamefont {V.}~\bibnamefont {Dunjko}},\ and\
  \bibinfo {author} {\bibfnamefont {M.}~\bibnamefont {Olshanii}},\ }\bibfield
  {title} {\bibinfo {title} {Thermalization and its mechanism for generic
  isolated quantum systems},\ }\href {https://doi.org/10.1038/nature06838}
  {\bibfield  {journal} {\bibinfo  {journal} {Nature}\ }\textbf {\bibinfo
  {volume} {452}},\ \bibinfo {pages} {854} (\bibinfo {year}
  {2008})}\BibitemShut {NoStop}%
\bibitem [{\citenamefont {Santos}\ and\ \citenamefont {Rigol}(2010)}]{Rigol10}%
  \BibitemOpen
  \bibfield  {author} {\bibinfo {author} {\bibfnamefont {L.~F.}\ \bibnamefont
  {Santos}}\ and\ \bibinfo {author} {\bibfnamefont {M.}~\bibnamefont {Rigol}},\
  }\bibfield  {title} {\bibinfo {title} {Onset of quantum chaos in
  one-dimensional bosonic and fermionic systems and its relation to
  thermalization},\ }\href {https://doi.org/10.1103/PhysRevE.81.036206}
  {\bibfield  {journal} {\bibinfo  {journal} {Phys. Rev. E}\ }\textbf {\bibinfo
  {volume} {81}},\ \bibinfo {pages} {036206} (\bibinfo {year}
  {2010})}\BibitemShut {NoStop}%
\bibitem [{\citenamefont {Tavora}\ \emph {et~al.}(2014)\citenamefont {Tavora},
  \citenamefont {Rosch},\ and\ \citenamefont {Mitra}}]{Tavora13}%
  \BibitemOpen
  \bibfield  {author} {\bibinfo {author} {\bibfnamefont {M.}~\bibnamefont
  {Tavora}}, \bibinfo {author} {\bibfnamefont {A.}~\bibnamefont {Rosch}},\ and\
  \bibinfo {author} {\bibfnamefont {A.}~\bibnamefont {Mitra}},\ }\bibfield
  {title} {\bibinfo {title} {Quench dynamics of one-dimensional interacting
  bosons in a disordered potential: Elastic dephasing and critical speeding-up
  of thermalization},\ }\href {https://doi.org/10.1103/PhysRevLett.113.010601}
  {\bibfield  {journal} {\bibinfo  {journal} {Phys. Rev. Lett.}\ }\textbf
  {\bibinfo {volume} {113}},\ \bibinfo {pages} {010601} (\bibinfo {year}
  {2014})}\BibitemShut {NoStop}%
\bibitem [{\citenamefont {Mitra}(2018)}]{AMreview}%
  \BibitemOpen
  \bibfield  {author} {\bibinfo {author} {\bibfnamefont {A.}~\bibnamefont
  {Mitra}},\ }\bibfield  {title} {\bibinfo {title} {Quantum quench dynamics},\
  }\href@noop {} {\bibfield  {journal} {\bibinfo  {journal} {Annual Review of
  Condensed Matter Physics}\ }\textbf {\bibinfo {volume} {9}},\ \bibinfo
  {pages} {245} (\bibinfo {year} {2018})}\BibitemShut {NoStop}%
\bibitem [{\citenamefont {Znidaric}(2020)}]{Znidaric20}%
  \BibitemOpen
  \bibfield  {author} {\bibinfo {author} {\bibfnamefont {M.}~\bibnamefont
  {Znidaric}},\ }\bibfield  {title} {\bibinfo {title} {Weak integrability
  breaking: Chaos with integrability signature in coherent diffusion},\ }\href
  {https://doi.org/10.1103/PhysRevLett.125.180605} {\bibfield  {journal}
  {\bibinfo  {journal} {Phys. Rev. Lett.}\ }\textbf {\bibinfo {volume} {125}},\
  \bibinfo {pages} {180605} (\bibinfo {year} {2020})}\BibitemShut {NoStop}%
\bibitem [{\citenamefont {Bulchandani}\ \emph {et~al.}(2022)\citenamefont
  {Bulchandani}, \citenamefont {Huse},\ and\ \citenamefont
  {Gopalakrishnan}}]{Bulchandani22}%
  \BibitemOpen
  \bibfield  {author} {\bibinfo {author} {\bibfnamefont {V.~B.}\ \bibnamefont
  {Bulchandani}}, \bibinfo {author} {\bibfnamefont {D.~A.}\ \bibnamefont
  {Huse}},\ and\ \bibinfo {author} {\bibfnamefont {S.}~\bibnamefont
  {Gopalakrishnan}},\ }\bibfield  {title} {\bibinfo {title} {Onset of many-body
  quantum chaos due to breaking integrability},\ }\href
  {https://doi.org/10.1103/PhysRevB.105.214308} {\bibfield  {journal} {\bibinfo
   {journal} {Phys. Rev. B}\ }\textbf {\bibinfo {volume} {105}},\ \bibinfo
  {pages} {214308} (\bibinfo {year} {2022})}\BibitemShut {NoStop}%
\bibitem [{\citenamefont {Altshuler}\ \emph {et~al.}(1997)\citenamefont
  {Altshuler}, \citenamefont {Gefen}, \citenamefont {Kamenev},\ and\
  \citenamefont {Levitov}}]{Altshuler1997}%
  \BibitemOpen
  \bibfield  {author} {\bibinfo {author} {\bibfnamefont {B.~L.}\ \bibnamefont
  {Altshuler}}, \bibinfo {author} {\bibfnamefont {Y.}~\bibnamefont {Gefen}},
  \bibinfo {author} {\bibfnamefont {A.}~\bibnamefont {Kamenev}},\ and\ \bibinfo
  {author} {\bibfnamefont {L.~S.}\ \bibnamefont {Levitov}},\ }\bibfield
  {title} {\bibinfo {title} {Quasiparticle lifetime in a finite system: A
  nonperturbative approach},\ }\href
  {https://doi.org/10.1103/PhysRevLett.78.2803} {\bibfield  {journal} {\bibinfo
   {journal} {Phys. Rev. Lett.}\ }\textbf {\bibinfo {volume} {78}},\ \bibinfo
  {pages} {2803} (\bibinfo {year} {1997})}\BibitemShut {NoStop}%
\bibitem [{\citenamefont {Abou-Chacra}\ \emph {et~al.}(1973)\citenamefont
  {Abou-Chacra}, \citenamefont {Thouless},\ and\ \citenamefont
  {Anderson}}]{Abou-Chacra_1973}%
  \BibitemOpen
  \bibfield  {author} {\bibinfo {author} {\bibfnamefont {R.}~\bibnamefont
  {Abou-Chacra}}, \bibinfo {author} {\bibfnamefont {D.~J.}\ \bibnamefont
  {Thouless}},\ and\ \bibinfo {author} {\bibfnamefont {P.~W.}\ \bibnamefont
  {Anderson}},\ }\bibfield  {title} {\bibinfo {title} {A selfconsistent theory
  of localization},\ }\href {https://doi.org/10.1088/0022-3719/6/10/009}
  {\bibfield  {journal} {\bibinfo  {journal} {Journal of Physics C: Solid State
  Physics}\ }\textbf {\bibinfo {volume} {6}},\ \bibinfo {pages} {1734}
  (\bibinfo {year} {1973})}\BibitemShut {NoStop}%
\bibitem [{\citenamefont {Gornyi}\ \emph {et~al.}(2016)\citenamefont {Gornyi},
  \citenamefont {Mirlin},\ and\ \citenamefont {Polyakov}}]{Gornyi2016}%
  \BibitemOpen
  \bibfield  {author} {\bibinfo {author} {\bibfnamefont {I.~V.}\ \bibnamefont
  {Gornyi}}, \bibinfo {author} {\bibfnamefont {A.~D.}\ \bibnamefont {Mirlin}},\
  and\ \bibinfo {author} {\bibfnamefont {D.~G.}\ \bibnamefont {Polyakov}},\
  }\bibfield  {title} {\bibinfo {title} {Many-body delocalization transition
  and relaxation in a quantum dot},\ }\href
  {https://doi.org/10.1103/PhysRevB.93.125419} {\bibfield  {journal} {\bibinfo
  {journal} {Phys. Rev. B}\ }\textbf {\bibinfo {volume} {93}},\ \bibinfo
  {pages} {125419} (\bibinfo {year} {2016})}\BibitemShut {NoStop}%
\bibitem [{\citenamefont {Monteiro}\ \emph {et~al.}(2021)\citenamefont
  {Monteiro}, \citenamefont {Tezuka}, \citenamefont {Altland}, \citenamefont
  {Huse},\ and\ \citenamefont {Micklitz}}]{Monteiro2021}%
  \BibitemOpen
  \bibfield  {author} {\bibinfo {author} {\bibfnamefont {F.}~\bibnamefont
  {Monteiro}}, \bibinfo {author} {\bibfnamefont {M.}~\bibnamefont {Tezuka}},
  \bibinfo {author} {\bibfnamefont {A.}~\bibnamefont {Altland}}, \bibinfo
  {author} {\bibfnamefont {D.~A.}\ \bibnamefont {Huse}},\ and\ \bibinfo
  {author} {\bibfnamefont {T.}~\bibnamefont {Micklitz}},\ }\bibfield  {title}
  {\bibinfo {title} {Quantum ergodicity in the many-body localization
  problem},\ }\href {https://doi.org/10.1103/PhysRevLett.127.030601} {\bibfield
   {journal} {\bibinfo  {journal} {Phys. Rev. Lett.}\ }\textbf {\bibinfo
  {volume} {127}},\ \bibinfo {pages} {030601} (\bibinfo {year}
  {2021})}\BibitemShut {NoStop}%
\end{thebibliography}

%
\end{document}